\documentclass{cta-author}
\usepackage{tabularx}
\usepackage{array} 
\usepackage{booktabs} 
\usepackage{stfloats}
\usepackage{threeparttable}

{}
{}
{}

\begin{document}

\supertitle{}

\title{AI-Based Crypto Tokens: The Illusion of Decentralized AI?}

\author{\au{Rischan Mafrur}}

\address{\add{}{Department of Applied Finance, Macquarie University, Australia}
\email{rischan.mafrur@mq.edu.au}}

\begin{abstract} The convergence of blockchain and artificial intelligence (AI) has led to the emergence of AI-based tokens, which are cryptographic assets designed to power decentralized AI platforms and services. This paper provides a comprehensive review of leading AI-token projects, examining their technical architectures, token utilities, consensus mechanisms, and underlying business models. We explore how these tokens operate across various blockchain ecosystems and assess the extent to which they offer value beyond traditional centralized AI services. Based on this assessment, our analysis identifies several core limitations. From a technical perspective, many platforms depend extensively on off-chain computation, exhibit limited capabilities for on-chain intelligence, and encounter significant scalability challenges. From a business perspective, many models appear to replicate centralized AI service structures, simply adding token-based payment and governance layers without delivering truly novel value. In light of these challenges, we also examine emerging developments that may shape the next phase of decentralized AI systems. These include approaches for on-chain verification of AI outputs, blockchain-enabled federated learning, and more robust incentive frameworks. Collectively, while emerging innovations offer pathways to strengthen decentralized AI ecosystems, significant gaps remain between the promises and the realities of current AI-token implementations. Our findings contribute to a growing body of research at the intersection of AI and blockchain, highlighting the need for critical evaluation and more grounded approaches as the field continues to evolve. \end{abstract}

\maketitle

\section{Introduction}\label{sec1}

Blockchain and artificial intelligence (AI) are two transformative technologies that are increasingly converging in innovative ways [\ref{ref:kersic2025}]. One notable outcome of this convergence is the emergence of AI-based tokens, which are digital assets designed to support decentralized platforms for AI computation, data sharing, and model deployment. These tokens aim to shift control over AI technologies away from centralized corporations, where users often lack meaningful ownership of their data [\ref{ref:defilippi2012}], and toward open, community-governed ecosystems. The core motivation behind these initiatives is to develop AI services that reflect the foundational principles of blockchain, i.e., decentralization, self-sovereignty, and user ownership over data and computational processes. These systems operate without centralized governance and are designed to avoid single points of failure while enhancing user privacy. By leveraging blockchain infrastructure, AI-token projects seek to promote transparency, traceability, and accessibility, while also introducing economic incentives that reward participation across the network. Examples include RENDER [\ref{ref:render2025}], AGIX (SingularityNET) [\ref{ref:goertzel2017}], OCEAN (Ocean Protocol) [\ref{ref:mcconaghy2022}], FET (Fetch.ai) [\ref{ref:weeks2018}], NMR (Numerai) [\ref{ref:craib2017}], and TAO (Bittensor) [\ref{ref:rao2020}], each enabling networks or marketplaces where AI models, datasets, or predictions are exchanged in a decentralized manner. 

The excitement surrounding AI-based tokens intensified following the release of ChatGPT in late 2022, which marked a pivotal moment in public engagement with generative AI. This breakthrough not only accelerated mainstream awareness of artificial intelligence, but also triggered a notable reaction in cryptocurrency markets. In particular, [\ref{ref:ante2024}, \ref{ref:saggu2023}] document that AI-related crypto assets experienced substantial abnormal returns in the immediate aftermath of ChatGPT’s launch, with peak gains exceeding 41\% within two weeks. Moreover, the majority of tokens in their sample exhibited significantly positive performance. These findings suggest that market participants responded strongly to the perceived thematic connection between emerging AI technologies and crypto assets associated with them. However, such rapid price appreciation raises important questions about the fundamental value of these tokens: do they represent genuine technological utility and decentralization, or are they merely an illusion of decentralization? Clarifying the extent to which AI-based tokens deliver substantive decentralization, rather than simply leveraging AI-related narratives for financial gain, is essential for assessing their long-term relevance and impact.

To address these concerns, this paper poses the following research questions aimed at providing a comprehensive investigation into the design, limitations, and future prospects of AI-based tokens:
\begin{enumerate}
    \item \textbf{Operation and Design:} How do current AI-based tokens function across different blockchain platforms, particularly in terms of their technical architecture, token utility, consensus mechanisms, and business models?

    \item \textbf{Limitations Compared to Centralized AI:} What are the primary limitations and challenges that prevent AI-token projects from offering clear advantages over existing centralized AI services?

    \item \textbf{Implementation Gaps:} What technical limitations currently hinder the development and broader adoption of AI-based tokens, particularly regarding off-chain computation, limited on-chain capabilities, and scalability challenges?

    \item \textbf{Future Directions:} What promising innovations and design strategies have the potential to support the next generation of AI-based tokens, particularly in enhancing practical utility, improving system sustainability, and fostering the development of more robust and inclusive decentralized AI ecosystems?

\end{enumerate}

To answer these questions, we begin by reviewing several leading AI-token projects and analyzing their underlying architectures. We then examine common challenges and structural weaknesses shared across these platforms, with a focus on their comparative disadvantages relative to centralized AI services. Finally, we explore emerging trends and innovations in both blockchain and AI that may help address current shortcomings. The aim is to identify viable paths forward for AI-token systems to evolve into more mature, impactful components of decentralized intelligence networks.

The remainder of this paper is structured as follows. Section~\ref{sec:defi_and_ai} outlines the current landscape of blockchain (especially decentralized finance) and artificial intelligence, highlighting their points of intersection. Section~\ref{sec:review_major_ai_token} presents a review of prominent AI-token projects, with a focus on their blockchain infrastructure, token utility, technical architecture, and underlying business models. Section~\ref{sec:challenges_limitations_ai_token} examines the key limitations and implementation challenges that hinder the effectiveness of existing systems. Finally, Section~\ref{sec:future_directions_ai_token} explores potential innovations and design strategies that could advance the development of AI-based tokens and support the emergence of more sustainable and impactful decentralized AI ecosystems.

\section{Related Work}
In this section, we review existing literature on the convergence of artificial intelligence and blockchain technologies, followed by a discussion on the illusion of decentralized AI and the speculative nature of crypto token economies.

\label{sec:defi_and_ai}

\subsection{The Convergence of Blockchain and AI}
Several papers have surveyed the integration of blockchain and AI across various domains. Some works focus on specific sectors, such as e-Health [\ref{ref:tagde2021}], food safety and quality control [\ref{ref:femimol2025}], supply chains [\ref{ref:charles2023}], traditional finance [\ref{ref:hosen2022}], and the metaverse [\ref{ref:yang2022}]. Other surveys provide broader overviews, examining the convergence of blockchain and AI across multiple industries [\ref{ref:salah2019}, \ref{ref:marwala2018}, \ref{ref:hussain2021}].

In contrast, this paper focuses specifically on a critical evaluation of well-known AI-based tokens, with an emphasis on their technical architectures, token utilities, and business models. We assess whether these tokens meaningfully advance the goals of decentralization or merely create the illusion of decentralization.
Since our focus is on crypto tokens, the closest related field is Decentralized Finance (DeFi). Therefore, in this section, we also review several current DeFi projects that incorporate AI technologies into their products. The intersection of Blockchain and AI represents a new frontier especially in the evolution of financial technologies, especially DeFi [\ref{ref:zetzsche2020}]. DeFi's built on blockchain infrastructures, removes the need for centralized intermediaries by enabling open, permissionless access to financial services [\ref{ref:michael2018}]. Meanwhile, AI technologies [\ref{ref:mccormick1985}], including machine learning [\ref{ref:zhou2021}], deep learning [\ref{ref:lecun2015}], and multi agent systems [\ref{ref:ferber1999}], enable data-driven decision-making, automation, and dynamic optimization. When combined, these technologies offer transformative potential: autonomous financial systems that are not only decentralized but also intelligent and adaptive.

DeFi protocols including decentralized exchanges (e.g., Uniswap [\ref{ref:adams2023}], Sushiswap [\ref{ref:sushi2025}]), lending platforms (e.g., Aave [\ref{ref:frangella2022}], Compound [\ref{ref:leshner2019}]), and stablecoin systems (e.g., MakerDAO and other algorithmic stablecoins [\ref{ref:ellinger2024}, \ref{ref:fiedler2023}]) are increasingly integrating artificial intelligence to improve operational efficiency and user experience. For example, research by Raun et al. [\ref{ref:raun2023}] investigates the behavior of MEV (Miner Extractable Value) bots operating within Flashbots’ private transaction channels [\ref{ref:weintraub2022}]. Their study demonstrates that machine learning models, trained on detailed transaction data, can predict successful bids in first-price MEV auctions with over 50\% accuracy. These models outperform conventional arbitrage strategies and highlight the effectiveness of adaptive constant bidding techniques in sandwich attacks [\ref{ref:wang2022}]. This work illustrates the growing relevance of AI in optimizing arbitrage mechanisms within DeFi markets and emphasizes its potential to enhance strategic decision-making under the unique constraints of blockchain-based financial systems.

In parallel with advancements in AI-driven arbitrage strategies, AI is also being deployed to enhance blockchain address attribution and transparency. A prominent example is Arkham Intelligence, a blockchain analytics platform that employs artificial intelligence to automatically label wallet addresses by analyzing transaction patterns and money flows. At the core of Arkham’s system is its proprietary engine, Ultra, which integrates on-chain and off-chain data to infer the identities of wallet holders and generate behavioral profiles for individuals and institutions. Its AI Entity Predictions feature further refines this process by applying machine learning models to predict address ownership with accompanying confidence scores, enabling scalable and semi-automated wallet classification. This approach not only supports forensic analysis and regulatory compliance but also represents a broader trend in leveraging AI to address the challenges of scale, accuracy, and interpretability in blockchain data analytics [\ref{ref:arkham2023}, \ref{ref:arkham2023ai}, \ref{ref:mafrur2025}].

What distinguishes the AI-based tokens discussed in this paper from the examples mentioned earlier is that their stated goal is not merely to apply AI within existing blockchain-based applications, but to decentralize both the data and the computational infrastructure underlying AI services. This vision challenges conventional assumptions regarding data governance, model ownership, and centralized institutional control. As noted in the systematic literature review by Keršič and Turkanović [\ref{ref:kersic2025}], decentralized artificial intelligence (DEAI) builds on the foundational principles of DeFi by envisioning systems in which AI models, agents, and datasets are registered, discovered, and executed through blockchain-based registries, decentralized marketplaces, and token-based incentive structures.

\subsection{Decentralization Illusions and Speculative Token Economies}

The term “illusion of decentralized AI” refers to cases in which blockchain-based AI projects present a decentralized architecture in theory, while retaining centralized control over core operations in practice. This perception is often maintained through structural features commonly associated with decentralization such as token-based governance, geographically distributed nodes, or autonomous project branding. Despite the fact that critical components like model training, data hosting, and protocol updates remain under the authority of a limited group of insiders. In such arrangements, decentralization exists in name only, as substantive control over decision-making, infrastructure, and economic flows remains concentrated among project founders, venture capital entities, or centralized service providers.

This decentralization illusion is not unique to AI-based tokens. Across the broader Web3 ecosystem, similar patterns have emerged where decentralization is assumed by design but fails in implementation. Researchers have increasingly emphasized the need for rigorous, quantifiable frameworks to assess whether authority, participation, and infrastructure are genuinely decentralized. In the absence of deliberate mechanisms to prevent concentration of control, blockchain-based systems risk replicating traditional power hierarchies while merely presenting an appearance of decentralization [\ref{ref:calzada2025}, \ref{ref:barbereau2023}]. In the context of decentralized AI, this means that tokenization or blockchain use alone is insufficient. Genuine decentralization must be evaluated based on who governs the models, how infrastructure is provisioned, and whether the community can meaningfully participate in development and oversight. The “illusion of decentralized AI” thus captures a critical gap between the ideals of Web3 and the realities of control in current AI-token ecosystems.

This conceptual gap is compounded by the fact that many so-called utility tokens have failed to deliver on their promised functionality, instead becoming vehicles for speculative trading. Empirical studies consistently show that the majority of utility tokens experience disproportionately high levels of speculation compared to actual platform usage [\ref{ref:zimmerman2020}, \ref{ref:silberholz2021}, \ref{ref:mei2022}]. A comprehensive analysis of 891 Ethereum-based tokens revealed a 90\% decline in average utility usage since 2017, even as speculative activity surged particularly following the rise of decentralized finance (DeFi), which introduced leveraged trading mechanisms that further shifted attention away from functional use [\ref{ref:silberholz2021}]. This trend spans multiple sectors, from decentralized storage to gaming, where token holders increasingly prefer to trade or hoard tokens in anticipation of price appreciation, rather than use them for access or services as originally intended. 

This disconnect is not merely behavioral but often embedded in token design. By making access tokens freely tradable, many projects inadvertently prioritize financial incentives over organic platform engagement. The Internet Computer Protocol (ICP) serves as a vivid example: despite initial hype and a multi-billion dollar valuation, its token collapsed by over 95\% shortly after launch, with real user adoption stagnating [\ref{ref:mei2022}]. Similar dynamics have been observed in Filecoin, where token value has been driven more by mining and speculation than by genuine demand for decentralized storage [\ref{ref:coinbase2023}]. These cases underscore a broader trend in tokenized ecosystems: despite ambitious narratives, many tokens have not fulfilled their intended utility roles, as speculative behavior systematically overrides platform use.

\begin{figure}[htbp]
    \centering
    \includegraphics[width=\columnwidth]{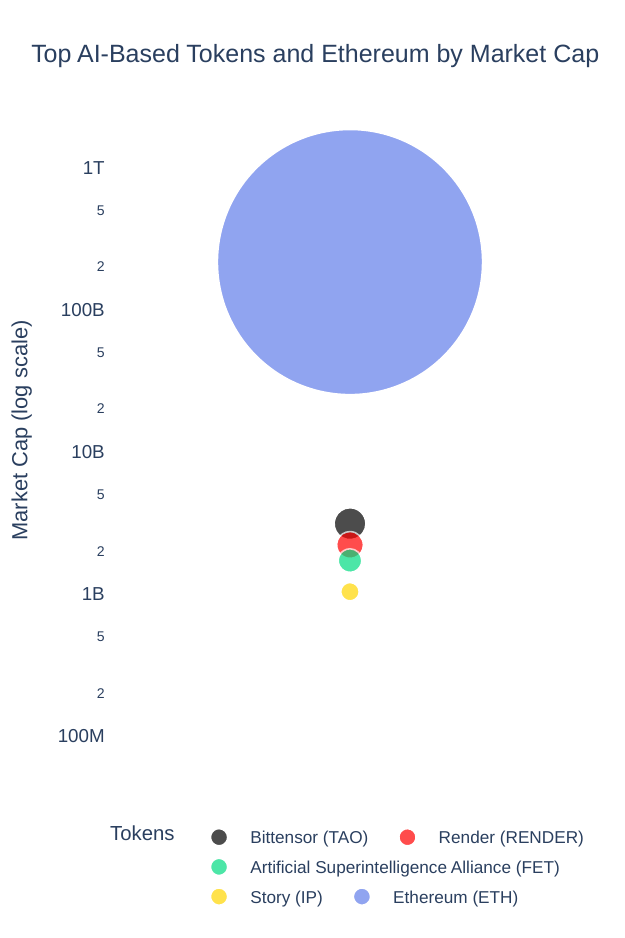}
    \caption{Market capitalization of selected AI-based tokens as of April 2025. The Artificial Superintelligence Alliance is currently operating under the FET ticker, with a planned transition to the ASI ticker following the merger of four AI-focused tokens: FET, AGIX, OCEAN, and CUDOS.}
    \label{fig:ai_token_marketcap}
\end{figure}

\section{Review of Major AI-Token Projects}
\label{sec:review_major_ai_token}
This section examines several leading projects that exemplify the AI-token paradigm, with a focus on their platform architecture, token functionality, consensus mechanisms, and business or application models, as summarized in Table~\ref{tab:ai_token_compared}. The selected projects are drawn from the top-ranked AI tokens by market capitalization, according to CoinMarketCap\footnote{https://coinmarketcap.com/view/ai-big-data/} and CoinGecko\footnote{https://www.coingecko.com/en/categories/artificial-intelligence}, and are limited to those that explicitly identify themselves as AI tokens. Figure~\ref{fig:ai_token_marketcap} presents the market capitalizations of the top AI-based tokens, with Ethereum included for comparative context. As shown, the top four AI tokens exhibit comparable valuations to one another; however, their market capitalizations remain significantly lower than that of Ethereum. This selection criterion ensures the analysis remains focused on initiatives that are AI-native by both design and narrative.

It is important to note that several prominent blockchain projects, although not originally developed as AI platforms, are now categorized under the AI sector on platforms such as CoinMarketCap and CoinGecko, reflecting recent strategic shifts toward AI applications. These include peer-to-peer storage networks such as IPFS, Arweave, and the Internet Computer Protocol (ICP) [\ref{ref:blythman2022}, \ref{ref:daniel2022}], which are increasingly used for decentralized AI data and model hosting. The Graph [\ref{ref:patelgraph}, \ref{ref:graph_ai_infra}], often referred to as the “Google of blockchain” for its decentralized indexing capabilities, has also rebranded its roadmap to support AI-related infrastructure, as outlined in its recent whitepaper “The Graph as AI Infrastructure.”. Chainlink [\ref{ref:chainlink2024}], a leading oracle network, is actively developing AI-driven oracle agents, and Layer 1 blockchain protocols such as Near [\ref{ref:near2025}] have launched dedicated initiatives aimed at supporting AI-native development. While these projects reflect the growing convergence between blockchain and AI, they are excluded from the core analysis in this paper, as they do not represent tokens that were initially or explicitly positioned as AI-focused. Their inclusion in the AI category, however, signals the expanding scope and appeal of decentralized AI as a frontier across the broader Web3 ecosystem. 

In this section, we review the top AI-based tokens include RENDER, a decentralized network for GPU computing [\ref{ref:render2025}]; Bittensor, a peer-to-peer platform for AI model collaboration [\ref{ref:rao2020}]; Fetch.ai, which enables autonomous AI agents for real-world coordination tasks [\ref{ref:weeks2018}]; SingularityNET, a decentralized marketplace for AI services [\ref{ref:goertzel2017}]; and Ocean Protocol, which facilitates decentralized data exchange for AI and analytics applications [\ref{ref:mcconaghy2022}], among others. 

{\scriptsize
\begin{table*}[!t]
\centering
\scriptsize 
\caption{Comparison of Major AI-Based Tokens}
\setlength{\tabcolsep}{3pt}
\renewcommand{\arraystretch}{1.2}
\begin{tabularx}{\textwidth}{|
  >{\raggedright\arraybackslash}p{0.9cm}|
  >{\raggedright\arraybackslash}p{2cm}|
  >{\raggedright\arraybackslash}p{2.2cm}|
  >{\raggedright\arraybackslash}p{2cm}|
  >{\raggedright\arraybackslash}p{3.3cm}|
  >{\raggedright\arraybackslash}p{1.8cm}|
  >{\raggedright\arraybackslash}X|}
\hline
\textbf{Token} & \textbf{Blockchain Base} & \textbf{AI Role} & \textbf{Computation} & \textbf{Token Utility} & \textbf{Consensus} & \textbf{Business Model} \\
\hline
FET   & Cosmos SDK & Agent coordination & Off-chain + Wasm & Payment, staking, governance & PoS (Tendermint) & Agent economy infra with token utility \\
RNDR  & Ethereum -> Solana     & GPU rendering infra & Off-chain        & Render fee, DAO voting       & PoS + PoH         & Burn-mint credits for compute jobs \\
TAO   & Substrate  & AI model marketplace & Off-chain        & Access, staking              & Proof of Intelligence & Rewards driven by model quality \\
OCEAN & Ethereum   & Data marketplace      & Off-chain        & Data access, staking, gov    & PoS (ETH)         & Monetization of training data via datatokens \\
AGIX  & Ethereum / Cardano & AI API marketplace & Off-chain   & API usage, governance        & PoS              & Commission on services, AGI ecosystem \\
CTXC  & Cortex Chain & On-chain inference   & On-chain (CVM)   & Gas, model usage royalties   & PoW (GPU)         & Miner-model reward loop, deterministic AI \\
NMR   & Ethereum   & Crowdsourced models   & Off-chain        & Stake-based reward for prediction quality & PoS (ETH) & Hedge fund powered by community models \\
IP & Custom multi-core chain (CometBFT / Cosmos SDK) & IP licensing and agent-based AI transactions & Primarily off-chain with onchain validation & Staking, royalties, licensing, fee payments & PoS (CometBFT) & Programmable IP marketplace for humans and AI agents \\
DBC   & Substrate  & GPU rental for AI     & Off-chain (GPU)  & Compute fees, validator staking & DPoS + PoI     & Distributed AI compute cloud with rewards \\
ALI   & Ethereum   & AI avatars / iNFTs    & Off-chain        & Avatar usage, NFT integration & PoS (ETH)        & Monetization of AI character IP \\
\hline
\end{tabularx}
\label{tab:ai_token_compared}
\end{table*}
}

\subsection{Render(RNDR)} 
Render [\ref{ref:render2025}] is a decentralized GPU computing network built on Ethereum [\ref{ref:buterin2013}], utilizing RENDER tokens (formerly branded as RNDR) to facilitate a marketplace for on-demand rendering services and AI-related computation. The platform connects users who require intensive GPU tasks, such as AI model training or 3D rendering, with a distributed pool of node operators offering spare GPU capacity. RENDER serves multiple functions within this ecosystem, acting as a payment token, staking asset, and governance instrument. Clients spend RENDER to access GPU power, while providers earn tokens by completing verified rendering jobs. The platform ensures task integrity through a Proof-of-Render validation mechanism, with job data recorded on-chain for transparency. RENDER's capped supply and staking-based job allocation create a competitive, incentive-aligned environment. Decentralized governance allows token holders to shape protocol upgrades and economic policies, including decisions such as expanding to other chains like Solana [\ref{ref:yakovenko2018}]. With a business model based on protocol fees and partnerships with content creation platforms, Render is emerging as a leading decentralized alternative to traditional cloud providers for GPU-based workloads.

\subsection{Bittensor (TAO)}
Bittensor [\ref{ref:rao2020}] is a decentralized AI network built on a custom blockchain using the Substrate framework [\ref{ref:py-substrate-interface}] and a novel consensus mechanism called Yuma, a Delegated Proof-of-Stake variant that ties block validation and token rewards to AI model performance. The network incentivizes contributions through its native token TAO, which is used for rewarding miners (AI model operators), paying transaction fees, staking, and participating in governance. Bittensor's architecture includes a main chain and specialized subnets, each focusing on distinct AI tasks such as natural language processing or image classification. Within these subnets, AI models are evaluated by peers and validators based on the relevance and quality of their responses, creating a reputation-based reward system. The platform’s Proof-of-Intelligence design shifts blockchain mining from computationally expensive hashing to the generation and validation of useful AI outputs. TAO functions as both the economic engine and governance tool, aligning incentives between AI quality and protocol development. Unlike centralized AI services, Bittensor offers an open and collaborative environment where models improve by interacting with each other, and decisions are made collectively by token holders. Its business model is centered around the utility and demand of the TAO token, with increasing adoption among researchers and developers drawn to its vision of a decentralized, community-governed AI infrastructure.

\subsection{Fetch.ai (FET)}
Fetch.ai [\ref{ref:weeks2018}] is a layer-1 blockchain platform built to support autonomous software agents and AI-driven applications, using a modified proof-of-stake consensus based on Cosmos SDK [\ref{ref:kwon2019}] and Tendermint [\ref{ref:buchman2016}]. Its architecture enables high-throughput, interoperable operations, supported by smart contracts and a framework for autonomous agents known as the Open Economic Framework. The native token, FET, facilitates transaction fees, agent services, staking, and on-chain governance. Agents within the network function as digital twins that negotiate, transact, and execute decisions based on real-time data and embedded AI models. FET is used not only for microtransactions among agents but also as collateral to prevent Sybil attacks and to unlock certain services. Fetch.ai integrates machine learning into agent behaviors, with off-chain model training and on-chain coordination, including experimental co-learning for federated model updates. The platform’s business model centers around fostering a decentralized agent economy, where value is driven by the utility of the FET token and the scale of AI-agent interactions. Use cases span from autonomous DeFi trading and supply chain optimization to smart city applications like parking and charging networks. With increasing adoption, an active developer community, and strategic partnerships, Fetch.ai represents a strong example of how blockchain infrastructure can coordinate decentralized AI systems in real-world settings.

\subsection{SingularityNET (AGIX)}

Recent research emphasizes the importance of breaking the AI oligopoly by fostering decentralized AI ecosystems. A distributed, blockchain-powered AI marketplace, such as SingularityNET has the potential to democratize AI development and access [\ref{ref:montes2019}, \ref{ref:kumar2021}]. SingularityNET [\ref{ref:goertzel2017}] is a decentralized platform designed to enable the creation, sharing, and monetization of AI services across a distributed network. Initially built on Ethereum [\ref{ref:buterin2013}], it has expanded to Cardano [\ref{ref:hoskinson2017}] and other chains to improve scalability and reduce transaction costs, reflecting a multi-chain strategy that enhances flexibility and resilience. The AGIX token serves multiple roles, including payment for AI services, staking for network incentives, and participation in governance decisions. Developers list their AI services on a blockchain-based registry, and payments are processed through a multi-party escrow system to ensure secure, trustless transactions. While AI models run off-chain, the blockchain coordinates service discovery, payment, and reputation tracking. SingularityNET also supports agent-to-agent interactions, where AI services can autonomously call and pay one another, forming an emerging machine-to-machine economy. Governance combines on-chain voting by AGIX holders with oversight from the SingularityNET Foundation, which leads strategic initiatives and community funding through programs like Deep Funding. The economic model links token value to network usage, aiming to create positive feedback between service growth, token demand, and developer participation. Adoption has been gradual yet steadily expanding, with applications spanning medical research, language translation, robotics, and decentralized finance (DeFi). Notable examples include Rejuve [\ref{ref:rejuve2023}], which integrates neural-symbolic AI and cross-organism omics data to address aging, and NuNet [\ref{ref:nunet2025}], which focuses on decentralized computing infrastructure. By enabling decentralized coordination and monetization of AI services, SingularityNET plays a central role in advancing blockchain-based AI marketplaces and offers valuable insights into the technical, economic, and governance challenges faced by decentralized AI ecosystems.

\subsection{Ocean Protocol}
Ocean Protocol [\ref{ref:mcconaghy2022}] is a decentralized data exchange infrastructure built on Ethereum, designed to facilitate secure and privacy-preserving sharing of data for AI and analytics. The platform uses smart contracts and datatokens (i.e., ERC-20 tokens representing access to specific datasets) to enable transactions, with compute-to-data mechanisms allowing AI models to be trained without exposing raw data. OCEAN, the native token, serves multiple purposes including payment for data services, staking for curation, participation in governance, and incentivizing data publication. Data providers can monetize their datasets while maintaining control, and curators can earn rewards by staking on valuable data. Ocean’s architecture combines on-chain asset management with off-chain compute orchestration, ensuring scalability and privacy. Governance is managed through OceanDAO, where token holders vote on grants and protocol changes, balancing technical upgrades with ethical considerations around data use. The business model supports both individual and institutional adoption, with growing traction in healthcare, automotive, and enterprise AI use cases. Ocean's open architecture allows community-run marketplaces and integration with DeFi, positioning it as a key infrastructure for the emerging data economy. By turning data into a tradable, permissioned asset, Ocean Protocol addresses major challenges in AI development (i.e., data access, privacy, and incentive alignment) which making it a central example of blockchain’s role in enabling decentralized AI ecosystems.

\subsection{Merging FET, AGIX, and OCEAN}
In a landmark move signaling the maturation of decentralized AI ecosystems, Fetch.ai, SingularityNET, and Ocean Protocol, which are three of the most prominent projects at the intersection of blockchain and artificial intelligence, announced their merger in 2024 to form the Artificial Superintelligence Alliance (ASI) [\ref{ref:asi2025}]. CUDOS, a comparatively smaller project focused on decentralized GPU computing, also joined the alliance. This strategic unification reflects a shared vision to accelerate the development of decentralized AI infrastructure and to challenge the increasing centralization of computational intelligence among a small number of dominant technology firms. The merger involved a phased integration of their respective tokens (i.e., FET, AGIX, and OCEAN) into a unified token called ASI, with defined conversion rates and a coordinated rollout across major blockchain platforms. The process began with the migration of AGIX and OCEAN into FET, followed by the introduction of ASI as a new token standard across exchanges and decentralized networks, effectively aligning the ecosystem under a single economic and governance structure. 

\begin{table}[h]
\centering
\scriptsize
\setlength{\tabcolsep}{4pt}
\renewcommand{\arraystretch}{1.2}
\begin{tabularx}{\linewidth}{|p{2.5cm}|p{2.5cm}|X|}
\hline
\textbf{Category} & \textbf{Token} & \textbf{Description} \\
\hline
AI Service Marketplaces & AGIX, TAO, IP & Marketplaces for buying and selling AI models and services. \\
AI Compute Networks & RNDR, DBC, GLM, RLC, CUDOS & Decentralized GPU and cloud compute infrastructure for AI tasks. \\
AI Data Marketplaces & OCEAN & Secure decentralized access to datasets for AI. \\
AI Agent/Automation & FET, NMR & Autonomous agents and crowdsourced AI coordination. \\
AI-Enhanced Web3app & ALI & AI-driven NFTs and digital avatars. \\
On-Chain AI Execution & CTXC & Smart contracts executing AI models on-chain. \\
\hline
\end{tabularx}
\caption{Classification of notable AI-based tokens in cryptocurrency ecosystems}
\label{tab:ai-token-classification}
\end{table}

Beyond those major AI-token platforms, several other projects contribute to the broader ecosystem by targeting specialized niches at the intersection of AI and blockchain. Numeraire (NMR) [\ref{ref:craib2017}] incentivizes accurate financial predictions through a stake-based competition model, turning algorithmic performance into a tradable asset. Story (IP) [\ref{ref:story2025}] introduces a peer-to-peer blockchain protocol for registering, exchanging, and monetizing intellectual property as programmable assets, enabling decentralized, automated licensing and collaboration across AI agents, software, and users through a multi-core architecture and cryptoeconomic incentives. Cortex (CTXC) [\ref{ref:chen2018}] pioneered on-chain AI inference by embedding machine learning capabilities within smart contracts, though it has seen limited adoption. Artificial Liquid Intelligence (ALI) [\ref{ref:aiprotocol2023}] powers Alethea AI's intelligent NFTs, combining generative AI with token-based governance to manage ethical content moderation. DeepBrain Chain (DBC) [\ref{ref:deepbrain2023}], one of the earliest attempts at decentralized AI compute, aimed to create a distributed GPU marketplace but struggled to scale against newer competitors.  Their diverse architectures and use cases provide valuable insights into how token economies can be structured to support different facets of decentralized AI, while also highlighting recurring challenges such as adoption barriers, scalability, ethical concerns, and the need for sustainable incentive mechanisms. To conclude this section, Table~\ref{tab:ai-token-classification} presents the classification of existing AI-based tokens within the cryptocurrency ecosystem.

\section{Challenges and Limitations of Current AI-Token Projects}
\label{sec:challenges_limitations_ai_token}
Although AI-based token projects are marked by innovation and ambition, they encounter a shared set of challenges and limitations. These difficulties often arise from the inherent mismatch between the intensive computational requirements of AI and the limitations of decentralized infrastructure, as well as from the relatively early stage of development compared to mature centralized AI platforms. This section examines the primary limitations that prevent AI-token ecosystems from delivering compelling advantages over their centralized counterparts. It also highlights critical implementation gaps observed across current projects. These issues directly relate to Research Questions 2 and 3, which address the reasons why existing AI-token platforms have not yet outperformed centralized services and what technical and architectural shortcomings remain unresolved in their current designs.

\subsection{Heavy Reliance on Off-Chain Computation}
A central limitation of current AI-token platforms is their dependence on off-chain infrastructure for executing AI tasks. Due to the computational intensity and data volume associated with AI workloads, blockchains (i.e., especially general-purpose platforms like Ethereum) are ill-suited for running machine learning models directly on-chain. As a result, most AI operations, including model inference and training, occur off-chain, with the blockchain serving primarily as a coordination and payment layer. In SingularityNET, for example, AI services are hosted on external servers, and the blockchain is used to manage service discovery and transactions. Ocean Protocol similarly stores data off-chain, using the blockchain only to manage metadata and access rights via datatokens. Fetch.ai’s autonomous agents operate off-chain, interacting with the chain primarily for settlement or identity verification. While this architecture is pragmatic, it introduces trust and transparency concerns. Users must rely on external actors to execute AI computations as claimed, without a robust mechanism for on-chain verification. This creates the risk of misreporting results or failing to perform services after receiving payment. Some projects mitigate this with escrow contracts, reputation systems, or delayed payment triggers, but these mechanisms add friction and cannot guarantee correctness. 

Projects like Cortex represent a rare effort to bring AI inference on-chain, enabling smart contracts to call AI models deployed directly within the blockchain environment. However, such approaches remain experimental and are largely confined to non-mainstream chains, with limited support across widely adopted ecosystems such as Ethereum. The inability to perform even basic inference on-chain restricts decentralized applications (dApps) from leveraging AI as a native component. This disconnect limits the development of autonomous agents and intelligent dApps capable of adapting their behavior based on real-time data analysis or environmental feedback.

Furthermore, the stateless nature of most smart contracts, where execution resets at each block [\ref{ref:bartoletti2025}] prevents on-chain learning or stateful AI behavior. While there have been theoretical proposals for “learning contracts” that evolve over time, practical implementations remain elusive. These would require persistent memory and continuous training mechanisms, possibly necessitating custom execution environments or specialized blockchains designed for stateful processing.

Consequently, most AI-token platforms rely on a hybrid model where the trustless guarantees of blockchain apply only to token transactions, while the core AI functionality remains off-chain and opaque. This undermines the vision of fully decentralized AI and raises the question of whether such systems offer meaningful advantages over centralized AI marketplaces. While blockchain infrastructure does enable benefits such as auditability, censorship resistance, and token-based incentives, the inability to verify AI computation on-chain remains a fundamental challenge to achieving true decentralization.

\subsection{Performance and Scalability Constraints}
AI-based token networks face significant scalability challenges across multiple dimensions. First, in terms of throughput and latency, blockchain infrastructure is not optimized for high-frequency, low-latency transactions. While commercial AI services may need to handle thousands of queries per second, public blockchains typically support far fewer transactions per second. This mismatch can bottleneck AI service delivery if every interaction requires an on-chain transaction. To address this, projects such as SingularityNET have implemented payment channels to bundle multiple service calls into a single on-chain settlement, and Fetch.ai has developed a custom high-performance chain to reduce latency for agent interactions. Nonetheless, the responsiveness and user experience of these decentralized systems often lag behind centralized AI APIs, especially for applications requiring rapid, high-volume data exchanges such as those in IoT environments.

Second, there are computational scaling limitations. The training and deployment of advanced AI models, particularly large language models, require substantial computational resources that decentralized networks currently cannot match. Centralized platforms operated by major technology firms (i.e., OpenAI [\ref{ref:openai2025}], AWS AI [\ref{ref:awsai2025}] , AI.Google [\ref{ref:googleai2025}]) can leverage vast clusters of GPUs and manage the large-scale coordination required for AI training. In contrast, decentralized networks encounter significant coordination overhead, latency from consensus mechanisms, and fragmentation of resources. Projects like Bittensor attempt to address this by distributing tasks across specialized subnets, but the total compute capacity and efficiency remain far below that of centralized systems. While federated learning has been proposed as a potential solution, real-world implementations in fully decentralized settings are still experimental and limited in scope.

Third, network scalability in terms of participant growth presents its own challenges. Many AI-token systems are constrained by the need for specialized hardware (as in Bittensor and Cortex) or advanced technical knowledge (as in Numerai), which limits participation. As networks grow, problems such as peer discovery, communication latency, and consensus efficiency become more pronounced. For example, Bittensor's model requires validators to evaluate miners' contributions, a task that becomes computationally intensive as the number of participants increases. Similarly, Fetch.ai’s agent ecosystem depends on efficient matchmaking and directory services to handle large-scale agent discovery.

In summary, decentralized AI networks are not yet capable of matching the speed, scale, and efficiency of centralized AI infrastructure. This performance gap poses a major constraint on their ability to serve high-demand, real-time AI applications, and remains a significant barrier to achieving parity with, or offering a clear advantage over, centralized AI solutions.

\subsection{Quality Control and Trust in AI Outputs}
Maintaining the quality and reliability of AI outputs presents a fundamental challenge in decentralized systems. In contrast to centralized providers, which typically ensure service quality through rigorous internal validation, service-level agreements, and responsive maintenance mechanisms, decentralized AI platforms often lack formal guarantees of correctness or performance. Some projects attempt to address this challenge through incentive-aligned token economics. Numerai requires data scientists to stake NMR tokens in support of their models, penalizing poor performance and rewarding accuracy. Bittensor similarly uses a competitive framework where miners are rewarded based on the perceived value of their AI contributions. These mechanisms align rewards with quality, but they also introduce risks such as validator collusion, model sabotage, or instability in participation due to financial losses. Designing these systems to be fair, resilient, and resistant to gaming remains an ongoing research challenge.

Another significant challenge lies in verifying the outputs of AI systems, particularly for complex or subjective tasks such as recommendation engines, financial forecasting (especially those relying on deep learning models), or natural language generation. While certain tasks such as image classification can be validated through majority voting among independent nodes, this approach is not scalable for most AI services. The so-called "verification dilemma" [\ref{ref:voneschenbach2021}, \ref{ref:jacovi2021}] arises when the correctness of an AI output cannot be easily or objectively confirmed without re-executing the computation, thereby undermining the benefits of distributed delegation. Proposals such as multi-provider consensus where compensation is issued only when multiple services produce consistent results show potential, but introduce additional cost and system complexity.

Moreover, trust in decentralized AI systems often falls back on off-chain mechanisms [\ref{ref:voneschenbach2021}, \ref{ref:jacovi2021}]. These include community reputation, off-chain reviews, or endorsement by recognized stakeholders. Ocean Protocol incorporates a staking-based curation system where users can stake OCEAN tokens on datasets they believe are valuable or trustworthy. While this provides an incentive to identify high-quality data, it remains vulnerable to bias, misinformation, and manipulation by early or coordinated actors.

Finally, the absence of centralized oversight raises concerns about harmful or unethical AI outputs. In a decentralized environment, there may be limited mechanisms for filtering biased, offensive, or malicious models. Without enforced community standards or governance protocols, such content may proliferate unchecked. Although some ecosystems, such as Alethea AI [\ref{ref:aiprotocol2023}], have implemented moderation councils or voting-based content governance, such models remain experimental and may be difficult to scale effectively.

In summary, decentralized AI platforms must contend with serious challenges in assuring service quality, verifying results, and safeguarding against misuse. Addressing these issues requires a combination of technical innovation, incentive design, and governance mechanisms that can function in open, trust-minimized environments. These concerns are central to understanding the current limitations of AI-token ecosystems and their prospects for delivering trustworthy and robust AI services at scale.

\subsection{Ecosystem Bootstrapping and Network Effects}
A major barrier to the success of AI-token platforms lies in the challenge of ecosystem bootstrapping. Unlike centralized AI services backed by large enterprises, decentralized platforms must cultivate both supply and demand in parallel, often without the benefit of existing user bases or capital resources. Many such projects face the classic two-sided market dilemma: AI developers are hesitant to contribute services or data without a substantial user base, while potential users see little reason to engage until the platform offers compelling, diverse, and trustworthy content. Projects like SingularityNET and Ocean Protocol have made significant strides in infrastructure and vision, yet the overall traction remains limited. While these platforms host a range of services and datasets, many offerings do not differ meaningfully from what is already available through conventional web APIs or public repositories (e.g., Kaggle [\ref{ref:kaggle2025}], Hugging Face [\ref{ref:huggingface2025}]). Transaction volumes and usage metrics remain modest, reflecting this chicken-and-egg dynamic.

A related challenge is competition with free or open alternatives. Much of the value proposition for AI-token platforms involves monetizing access to AI models or datasets. However, many datasets are already openly available, and leading AI models are frequently published as open source (e.g., Kaggle [\ref{ref:kaggle2025}], Hugging Face [\ref{ref:huggingface2025}]). Without a strong differentiator such as guaranteed privacy, novel data aggregation, or decentralized model collaboration, users may not be inclined to pay for services that can be accessed freely elsewhere. Ocean Protocol’s compute-to-data mechanism attempts to address this by enabling privacy-preserving AI computation, but widespread adoption of such features takes time and education.

Centralized incumbents present another formidable obstacle. Established cloud providers like AWS AI [\ref{ref:awsai2025}], AI.Google [\ref{ref:googleai2025}], and Microsoft Azure AI [\ref{ref:azureai2025}] offer a broad suite of integrated AI services with high reliability, scalability, and user-friendly interfaces. These platforms also benefit from free-tier offerings and enterprise-grade support, making them attractive for both small developers and large organizations. Decentralized networks, in contrast, often lack the economies of scale to compete on price. Due to crypto-economic overheads, redundancy, and infrastructural inefficiencies, blockchain-based AI services may be significantly more expensive [\ref{ref:capponi2024}, \ref{ref:wang2022}]. This makes it difficult to position them as cost-effective alternatives unless they offer distinct advantages in trust, privacy, or decentralization.

User experience also remains a critical limitation. Many AI-token ecosystems are designed with a developer-centric or crypto-native user base in mind, requiring users to manage digital wallets, acquire tokens, and understand blockchain mechanics [\ref{ref:voskobojnikov2021}]. This complexity presents a substantial barrier to mainstream adoption, particularly for businesses or non-technical users seeking seamless AI services. Although some projects have started addressing this gap through improved interfaces and embedded assistants, overall usability remains far behind that of traditional cloud platforms.

In summary, overcoming ecosystem inertia and achieving meaningful network effects remains one of the most difficult tasks facing AI-token projects. Without a compelling value proposition that clearly surpasses or complements existing centralized services, and without substantial efforts to simplify access and improve incentives for early participants, many decentralized AI platforms risk stagnation. These ecosystem-level constraints are central to understanding the slow pace of adoption and the difficulty in delivering value beyond existing AI paradigms.

\subsection{Governance and Project Evolution Challenges}
Governance presents a significant challenge for decentralized AI networks, particularly in the context of continuous technological change, evolving economic models, and increasing regulatory scrutiny. Unlike centralized organizations that can rapidly adapt and implement strategic changes, decentralized platforms must rely on token-holder consensus, which can introduce delays, political friction, and coordination difficulties (e.g., Ethereum (ETH) vs. Ethereum Classic (ETC) [\ref{ref:kiffer2017}], Bitcoin (BTC) vs. Bicoin Cash (BCH)[\ref{ref:kwon2019bitcoin}]).

Protocol upgrades are essential for ensuring the long-term viability of decentralized AI systems, particularly as both artificial intelligence and blockchain technologies continue to evolve at a rapid pace. Projects must remain adaptable, incorporating advances in machine learning, transitioning to more scalable blockchain infrastructures, and updating interoperability frameworks. However, decentralized governance can introduce significant delays in executing such changes. For instance, transitioning to a multi-chain architecture often requires complex coordination across token bridges and swap mechanisms. While some initiatives benefit from the guidance and coordination of a central foundation facilitating relatively efficient implementation fully decentralized projects without such leadership may face challenges in achieving consensus and executing upgrades in a timely manner.

Many AI-token ecosystems operate under intricate incentive structures involving staking rewards, performance-based payouts, and token inflation mechanisms. These parameters often require iterative calibration based on real-world user behavior and participation trends. Poorly tuned parameters such as overly punitive staking losses or skewed reward distributions can deter contributors and destabilize the network. Projects like Numerai and Bittensor have demonstrated the importance of aligning incentives carefully, but adjusting these systems in a decentralized and transparent manner remains challenging. Governance decisions ideally should be data-driven, yet in the absence of centralized analytics and feedback loops, communities often lack the tooling or coordination needed for such refinement.

Regulatory uncertainty further complicates governance [\ref{ref:clarke2019}, \ref{ref:uzougbo2024}]. AI-token platforms operate at the intersection of two rapidly evolving and heavily scrutinized domains: cryptocurrency and artificial intelligence. The legal classification of utility tokens remains ambiguous in many jurisdictions, and the content produced by decentralized AI systems (e.g., ranging from predictive analytics to generative media) may raise questions of liability, compliance, and content moderation. For instance, Ocean Protocol must tread carefully in handling sensitive datasets to avoid violating privacy laws such as GDPR. Similarly, generative AI networks may need to restrict or filter content to avoid regulatory backlash. In centralized systems, compliance can be managed through internal policies and legal oversight. In decentralized settings, enforcing content standards or responding to legal demands becomes more difficult, especially in the absence of a formal entity or accountable governance structure. This legal ambiguity can deter institutional participation, limit platform capabilities, or lead to fragmentation as forks emerge to implement varying levels of control.

In summary, decentralized AI ecosystems face a delicate balance between openness and adaptability. Without robust governance frameworks capable of responding to technical, economic, and legal pressures, these projects risk stagnation or fragmentation. Developing more agile, data-informed, and transparent governance mechanisms is therefore critical to ensuring their long-term evolution and resilience.

\subsection{Accountability and Legal Responsibility in Decentralized AI}

While decentralized AI-token projects promise openness, autonomy, and trustless execution, they also raise pressing concerns around accountability and legal responsibility. Traditional AI systems typically operate under a central provider who can be held liable for malfunctions or harm. In contrast, decentralized platforms like Fetch.ai, Ocean Protocol, and SingularityNET distribute control across pseudonymous actors, making it difficult to determine who, if anyone, bears responsibility in the event of failure or misuse. As recent legal scholarship points out, these systems often create a “liability vacuum,” where no single party is clearly accountable under existing legal frameworks \footnote{https://www.blockchainandthelaw.com/2023/04/dao-deemed-general-partnership-in-negligence-suit-over-crypto-hack-prompting-decentralized-orgs-to-rethink-corporate-formation/}.

Several high-profile projects attempt to mitigate this gap through hybrid governance models or foundation structures. For instance, Ocean Protocol and Fetch.ai maintain formal legal entities that can serve as regulatory touchpoints, while projects like Render and Bittensor rely on DAO-based voting mechanisms for internal oversight. However, token-based governance does not imply legal personhood, and scholars have raised concerns that DAO participants or token holders could, under certain conditions, be deemed members of unincorporated partnerships, thereby exposing them to liability. This has been exemplified by the U.S. court ruling in the Ooki DAO case\footnote{https://www.cftc.gov/PressRoom/PressReleases/8715-23}, which signaled that decentralized participation may not exempt individuals from legal accountability.

Technical solutions such as proof-of-quality staking mechanisms (e.g., Numeraire, Bittensor) or auditability through blockchain-based provenance have been proposed to enhance internal accountability. Yet these remain voluntary and do not fulfill the legal requirements of redress, especially under emerging global AI regulations. The EU AI Act [\ref{ref:marshall2024}], for example, imposes obligations such as transparency, data governance, and post-deployment monitoring responsibilities that decentralized AI networks may struggle to assign without a designated operator.

As legal and regulatory frameworks evolve, AI-token ecosystems must address the mismatch between distributed operation and centralized legal responsibility. Without clearer outline of liability whether through DAO registration, delegated entities, or protocol-level insurance, projects risk undermining user trust and exposing contributors to unforeseen legal risks. Accountability in decentralized AI will require more than on-chain governance; it demands structural innovations that bridge blockchain design with real-world legal norms.

\subsection{Delivering Value Beyond Centralized AI}

While this study focuses on AI-based tokens, similar challenges have emerged across a broader range of blockchain projects, suggesting that the underperformance of utility tokens is often a systemic issue. During the 2017–2018 ICO boom, numerous projects across sectors such as IoT, cloud storage, and social media launched tokens with ambitious promises but failed to achieve meaningful user adoption. Empirical studies [\ref{ref:silberholz2021}, \ref{ref:zimmerman2020},  \ref{ref:mei2022}] show that token valuations frequently decouple from actual usage, indicating that speculative trading often outweighs real utility. Helium, for example, attracted substantial investment and deployed widespread infrastructure, yet by 2022, its IoT data network generated only minimal revenue, highlighting a disconnect between token-driven supply growth and end-user demand \footnote{https://cointelegraph.com/news/critique-on-helium-s-6-5k-monthly-revenue-causes-a-stir}.

A similar pattern is evident in Filecoin, which built massive decentralized storage capacity but saw minimal utilization in its early years, with just 3.8\% of space filled compared to 40–70\% for centralized providers [\ref{ref:coinbase2023}]. Despite incentives and subsequent efforts to boost demand, adoption remained slow. These cases reveal a common issue in tokenized ecosystems: while token rewards can effectively stimulate supply, generating sustained demand for decentralized services, particularly when competing with established centralized alternatives, proves far more difficult. Such outcomes underscore that the limitations observed in AI-token projects are not unique, but reflective of broader structural challenges in aligning token models with real-world utility. These examples from outside the AI domain illustrate that token underperformance is frequently a matter of lack of product-market fit and network effects, rather than something intrinsic to AI.

Given the challenges outlined in Sec.~\ref{sec:challenges_limitations_ai_token}, it is understandable why current AI-token ecosystems have struggled to deliver value that clearly surpasses centralized AI alternatives. At present, many decentralized solutions involve greater complexity, slower performance, and limited usability. As a result, the theoretical benefits of decentralization such as censorship resistance, transparent governance, and collective ownership have not yet proven compelling enough to drive mass adoption outside of ideologically motivated communities or early adopters.

A central promise of decentralized AI is the potential to unlock network effects that centralized platforms cannot replicate. In theory, a decentralized model could aggregate contributions from thousands of independent developers, data providers, and AI agents, creating a richer and more diverse ecosystem than any single company could manage. Projects like Numerai, which crowdsources predictive models from a global community of data scientists, and Bittensor, which aims to build an open, collectively owned neural network, are early examples of this potential. However, in other domains, such as AI model hosting or dataset sharing, centralized platforms like Hugging Face [\ref{ref:huggingface2025}] and Kaggle [\ref{ref:kaggle2025}] currently maintain clear advantages in both scale and user engagement.

Importantly, surpassing centralized AI also requires doing what centralized systems cannot or will not do. Privacy-preserving AI training, as envisioned by Ocean Protocol’s compute-to-data framework, offers one such frontier. If successfully implemented, it could enable secure AI applications on sensitive datasets (e.g., medical records) that are currently inaccessible to cloud-based AI platforms. Similarly, decentralized ownership of foundational AI models, as proposed by Bittensor, could introduce new paradigms in collective training and governance of large-scale AI systems. These scenarios represent promising opportunities to offer differentiated, rather than merely equivalent, value relative to centralized incumbents. Table~\ref{tab:centralized_vs_decentralized_ai} presents a comparison and analogy between AI-based tokens and existing centralized AI services.

\begin{table*}[!b]
\centering
\begingroup
\scriptsize
\caption{Centralized vs Decentralized AI services}
\setlength{\tabcolsep}{3pt}
\renewcommand{\arraystretch}{1.2}
\begin{tabularx}{\textwidth}{|
  >{\raggedright\arraybackslash}p{2.5cm}|
  >{\raggedright\arraybackslash}p{5.2cm}|
  >{\raggedright\arraybackslash}p{4.9cm}|
  >{\raggedright\arraybackslash}X|}
\hline
\textbf{Category} & \textbf{Centralized AI Services} & \textbf{Decentralized AI Tokens} & \textbf{Notes / Analogy} \\
\hline
Platform Example & OpenAI, AWS SageMaker, Google Colab, Hugging Face, Kaggle & FET, AGIX, RNDR, TAO, OCEAN, NMR & Centralized SaaS vs Tokenized infra \\
Compute Execution & On centralized servers (e.g., AWS, MS Azure) & Mostly off-chain via node operators & Same architecture with trust layer shift \\
Model Hosting & Proprietary cloud-based APIs & Hosted by community nodes or contributors & Still lacks full transparency \\
Access Control & Accounts + API keys & Token-based access via marketplaces & Anyone with tokens can use services \\
Monetization & Credit card, subscription & Token-based payments, staking, royalties & Aligned incentives for contributors \\
Governance & Centralized company policies & DAO governance via token voting & In theory more democratic \\
Transparency & Limited (black-box models) & Varies: some on-chain traces, some off-chain & Verification is still weak \\
Incentive Design & Users pay; few earn & Nodes earn via task rewards & Participatory model \\
Data Control & Company-curated, licensed datasets, open datasets & Open or user-contributed datasets & Ocean Protocol leads here \\
Agent Interaction & Pre-scripted APIs or assistants & Agents act autonomously (e.g., FET) & Inspired by MAS (Multi-Agent Systems) \\
Marketplace Model & App store-style (central) & P2P service exchange (e.g., RNDR) & Token as medium of exchange \\
Compute Verification & Trust the provider & Some use staking or zk proofs & Still experimental \\
\hline
\end{tabularx}
\label{tab:centralized_vs_decentralized_ai}
\endgroup
\end{table*}

In summary, while AI-token projects face significant limitations including high overhead, limited scalability, quality assurance challenges, and underdeveloped network effects there are unique opportunities where decentralization may ultimately provide a compelling advantage. The current impact of these projects remains limited, but ongoing innovation in incentive design, on-chain verifiability, privacy-preserving computation, and user experience continues to move the field forward. In the following section, we examine the emerging solutions and research directions that may help overcome these barriers and enable decentralized AI ecosystems to realize their full potential.

\begin{table*}[ht]
\scriptsize
\setlength{\tabcolsep}{4pt}
\renewcommand{\arraystretch}{1.2}
\centering
\begin{threeparttable}

\caption{Summary of Key Achievements and Shortcomings of Selected AI Token Projects}
\begin{tabularx}{\textwidth}{|l|X|X|}
\hline
\textbf{Project Token} & \textbf{Key Achievements} & \textbf{Notable Shortcomings} \\
\hline
FET (Fetch.ai) & 
Launched a Cosmos-based mainnet enabling agent-based coordination and WASM smart contracts. &
Most AI tasks run off-chain. Governance is still significantly influenced by the foundation, raising centralization concerns. \\
\hline
RNDR (Render)& 
Enabled a decentralized GPU rendering network with measurable growth (e.g., 121\% QoQ usage increase in Q2 2023)\tnote{a}. Adopted a burn-and-mint model and DAO-based community governance. &
Most rendering tasks are conducted off-chain, requiring trust in node integrity. Initial development was centralized via OTOY\tnote{b}, and current focus remains narrow (e.g., graphics rendering). \\
\hline
TAO (Bittensor)& 
Pioneered mechanism where model quality determines block rewards [\ref{ref:rao2020}]. Established a peer-evaluated AI model marketplace and introduced subnet specialization. &
Scalability remains a challenge, as verifying model quality requires peer coordination. The outputs are not state-of-the-art, and participation barriers (hardware, ML expertise) limit decentralization. \\
\hline
OCEAN (Ocean Protocol) & 
Introduced Compute-to-Data\tnote{c}, allowing algorithms to run on private datasets without revealing raw data. Enabled data tokenization for monetization and supported DeSci integrations. &
The marketplace sees limited real demand. Many datasets are test entries or unused, and competition from free, centralized alternatives (e.g., Kaggle) undercuts its utility proposition. \\
\hline
AGIX (SingularityNET) & 
Built a decentralized AI service marketplace and led the ASI Alliance merger (AGIX, FET, OCEAN). Continued to fund AGI-related research (e.g., OpenCog Hyperon). &
Platform usage is low, with many listed services seeing no traffic. Trust mechanisms for service quality are lacking, and multi-chain operations introduce technical complexity. \\
\hline
CTXC (Cortex) & 
Implemented on-chain model inference via the Cortex Virtual Machine (CVM) with GPU-backed PoW consensus. &
Despite the novel architecture, the project has seen minimal developer engagement, and scaling inference fully on-chain remains infeasible with current resources. \\
\hline
NMR (Numeraire) & 
Enabled crowdsourced model predictions for a hedge fund, using stake-weighted incentives to reward accuracy. &
The platform is heavily centralized in its model selection process, and broader adoption beyond the Numerai ecosystem is limited. \\
\hline
IP (Story) & 
Proposed a blockchain-based licensing system for AI-generated and co-created IP, enabling programmable royalties. &
The ecosystem remains underdeveloped, with little visibility or user adoption, and limited clarity on execution scalability. \\
\hline
DBC (DeepBrain Chain) & 
Offered a decentralized GPU marketplace for AI compute, utilizing a Substrate chain and validator staking. &
The project has suffered from poor documentation and declining community activity. Verifiability of output remains unsolved. \\
\hline
ALI (Alethea AI) & 
Pioneered interactive NFTs powered by AI personas and showcased avatar monetization via AI dialogue systems. &
The use case is narrow and largely speculative, with limited mainstream demand or technical innovation in AI processing. \\
\hline
\label{tab:summary_achievement_shortcomings}

\end{tabularx}
\begin{tablenotes}
\scriptsize
\item[a] https://www.coinfeeds.ai/crypto-blog/render-network
\item[b] https://home.otoy.com/the-company/
\item[c] https://docs.oceanprotocol.com/developers/compute-to-data
\end{tablenotes}
\end{threeparttable}
\end{table*}

\section{Future Directions and Opportunities}
\label{sec:future_directions_ai_token}
Despite current limitations, the intersection of AI and blockchain continues to evolve, with several promising developments on the horizon. Advances in on-chain verifiability (e.g., zero-knowledge proofs such as zkML [\ref{ref:chen2024zkml}], zkPoT [\ref{ref:abbaszadeh2024}], [\ref{ref:kersic2024}, \ref{ref:xing2025}]), improved incentive systems [\ref{ref:jaiman2022}], privacy-preserving computation, cross-chain interoperability, composability, and modular blockchain architectures are paving the way for more trustless and scalable AI networks. Supported by active research and experimentation, these developments represent meaningful steps toward building open, intelligent, and community-driven AI infrastructures that can deliver value beyond what centralized systems are capable of.

\subsection{Verifiable Off-Chain Computation and AI Oracles}
A major step toward trustless decentralized AI is enabling on-chain verification of off-chain computation. Zero-knowledge proofs for machine learning (zkML [\ref{ref:chen2024zkml}]), for deep learning (zkPoT [\ref{ref:abbaszadeh2024}]) allow an AI provider to submit not only an output but also a cryptographic proof that the output was generated by a specific model on a given input. Although currently limited to lightweight models, ongoing research and hardware acceleration may soon make this practical for broader AI use. Trusted Execution Environments (TEEs) [\ref{ref:sabt2015}] provide an alternative, enabling secure model execution within hardware enclaves and generating attestations that can be verified on-chain. This shifts trust from server operators to hardware providers and is already being explored in federated learning contexts. Additionally, AI oracles, such as those under development by Chainlink [\ref{ref:breidenbach2021}, \ref{ref:chainlink2024}], propose a model in which multiple nodes independently execute an AI task and report only when a consensus is reached. This quorum-based approach could make AI inference verifiable and composable within smart contracts, paving the way for trust-minimized AI services accessible on-chain.

\subsection{Specialized Blockchains and Layer-2s for AI}
Emerging infrastructure tailored for AI workloads may address current performance bottlenecks in decentralized networks. AI-dedicated Layer-1s could incorporate native GPU support, larger block sizes, and sharding by AI task type (e.g., language or vision), enabling scalable, parallelized inference and training. Novel consensus mechanisms like Proof-of-Useful-Work (PoUW) [\ref{ref:lihu2020}, \ref{ref:ball2017}], where miners perform machine learning computations, could further align network security with productive AI contributions, though this depends on solving verification challenges. Academic proposals like Coin.AI [\ref{ref:baldominos2019}] have outlined mechanisms where block validity depends on a trained model meeting performance thresholds, with verification handled efficiently by the network. A real-world example is Bittensor which rewards miners based on the value their models contribute to a shared AI ecosystem. If implemented securely, PoUW could align blockchain incentives with the generation of valuable AI artifacts, transforming blockchain networks into decentralized compute resources. 

Alternatively, Layer-2 solutions [\ref{ref:huang2024}, \ref{ref:song2024}, \ref{ref:mafrur2025}, \ref{ref:tortola2024}, \ref{ref:sguanci2021}], for instance, Golem (GLM) [\ref{ref:zawistowski2016}] iExec (RLC) [\ref{ref:iexec2025}], which focuses on decentralized GPU computing, has migrated from Ethereum Layer 1 to Ethereum’s Layer 2. In partucular, zk-rollups, offer a promising path. By batching off-chain AI model executions and submitting zero-knowledge proofs to Layer-1s, rollups can significantly reduce costs and enable verifiable inference at scale. Projects like ZkMatrix [\ref{ref:cong2024}] hint at this direction. 

\subsection{Modular Blockchain Architectures for Decentralized AI}
A key architectural shift shaping the future of decentralized AI is the transition from monolithic to modular blockchain designs. Traditional monolithic blockchains (e.g., Bitcoin, Ethereum (pre-rollups)) require each node to perform all core functions consensus, data availability, and transaction execution which limits scalability and makes it difficult to accommodate computation heavy tasks such as AI model training or inference. In this design, every node must process every transaction, creating performance bottlenecks that hinder the integration of advanced AI workloads. Modular blockchains (e.g., Cosmos, Celestia, Ethereum (with rollups)), by contrast, decouple these functions into separate layers or components. For example, one layer might focus solely on transaction ordering and consensus, while a separate execution layer handles the state changes and computational logic. This division of labor allows each component to be optimized independently, improving efficiency and throughput while preserving system wide coherence.

For decentralized AI, the implications of modularity are particularly significant. AI applications often require flexible, high-throughput environments to support data processing, model execution, and real-time responsiveness. Modular architectures enable the deployment of specialized execution layers tailored to AI-specific tasks, without placing additional strain on the base chain. This model not only enhances scalability but also encourages composability where different components of the AI stack (e.g., data access, computation, governance) can operate on separate yet interoperable chains. Projects such as the Artificial Superintelligence Alliance, which includes Fetch.ai (FET) and SingularityNET (AGIX), illustrate this direction by envisioning interoperable AI modules that can operate across a network of purpose built chains. Rather than waiting for a single blockchain to meet all computational and coordination needs, modularity allows AI developers to build scalable systems by composing specialized layers, each responsible for a distinct function. This paradigm shift represents a more adaptable and sustainable path forward for decentralized AI infrastructure.

\subsection{Federated Learning and Collaborative Model Training}
Future decentralized AI systems may transition from merely deploying models to collaboratively building them. Blockchain-enabled federated learning [\ref{ref:qu2022}, \ref{ref:qu2020}, \ref{ref:toyoda2020}] allows participants to contribute to the training of a shared model without the need to exchange raw data. Smart contracts can facilitate training rounds, verify updates, and distribute token rewards based on the quality of contributions, which is measured by improvements in model performance. 

For example, [\ref{ref:ren2024}] propose FLCoin, a federated learning architecture with a two-layer blockchain: a local FL layer selects committee nodes, and an outer blockchain layer records model updates. Their experiments show stable scalability and reduced communication overhead as the number of nodes increases. Similarly, [\ref{ref:shayan2021}] introduce Biscotti, a P2P federated learning system where blockchain and homomorphic encryption coordinate multi-party stochastic gradient descent (SGD). This fully decentralized approach enforces honest aggregation, though it comes with additional consensus overhead. 
[\ref{ref:mothukuri2021}] propose FabricFL, which integrates Hyperledger Fabric with federated learning. Their framework combines FL updates with blockchain verification, ensuring that each model update is immutably logged and auditable. By using an enterprise blockchain, FabricFL introduces verifiable tracing into each training round, deterring malicious clients and enabling transparent accountability within decentralized AI systems.
This structure not only incentivizes honest participation but also paves the way for decentralized, privacy-preserving model development in sectors such as healthcare and finance.

Data cooperatives or data DAOs [\ref{ref:potts2021}, \ref{ref:thomas2024}, \ref{ref:petreski2024}, \ref{ref:zichichi2022}] may also emerge, enabling communities to pool data, vote on training goals, and share model revenue proportionally via on-chain governance. Additionally, incentivized AI challenges, inspired by projects like Numerai [\ref{ref:craib2017}, \ref{ref:numerai2025}], could be generalized to other domains, creating token-driven competitions for predictive modeling across areas such as climate, finance, or supply chains. Together, these approaches offer a blueprint for collective AI development with transparent incentives and governance.

\subsection{Enhanced Tokenomics and Governance for Sustainability}
Sustainable AI-token ecosystems will likely require more sophisticated economic and governance frameworks. Dynamic incentive mechanisms, such as utilizing Data Shapley Value [\ref{ref:nguyen2021}], could better align contributions with quality and rewarding consistent performance. Concepts like proof-of-quality [\ref{ref:zhang2024}, \ref{ref:ayaz2020}], where outputs are backed by slashed collateral, may further improve accountability. To promote long-term viability, value capture mechanisms such as fee redistribution, token burning, and protocol revenue sharing can be employed to complement utility-driven demand. This approach is exemplified by the treasury management strategies implemented by Ocean Protocol [\ref{ref:ocean_token_model}]. Lastly, hybrid governance models that combine token voting with expert input (e.g., advisory councils or merit-based influence) could support ethical oversight and technically informed decision-making, ensuring that decentralized AI evolves responsibly and inclusively.

\subsection{Real-World Integration}
To achieve real-world utility, AI-token networks will likely need to adopt hybrid architectures that integrate seamlessly with existing enterprise systems while adapting to regulatory and infrastructural realities. Blockchain technology has the potential to secure and coordinate deep learning processes in decentralized environments. For instance, [\ref{ref:shafay2023}] provides a comprehensive survey of blockchain-based deep learning frameworks, categorizing them by consensus mechanisms, privacy considerations, and application domains. The authors highlight how blockchain can ensure model integrity and facilitate fair computation in distributed training processes. Similarly, [\ref{ref:das2024}] propose a blockchain-encrypted deep learning system for managing healthcare data. Their BcEs-DLM scheme leverages blockchain smart contracts to handle encrypted medical records and regulate access to the models. Another example is DeepChain [\ref{ref:weng2019}], a framework that combines federated deep learning with blockchain-driven rewards. In DeepChain, each participant's local training contributions are logged on-chain, with blockchain-based token rewards encouraging honest behavior. This protocol-level integration ensures complete auditability, with every model update verifiably recorded, thus deterring malicious actors while preserving data privacy. The DeepChain prototype demonstrates that combining blockchain with secure deep neural network training can foster a fair and transparent collaborative learning environment. These studies demonstrate how integrating blockchain with deep learning can enhance security and foster trust in enterprise systems, particularly in sectors like healthcare.

\subsection{Cross-Sector Innovation Between AI Tokens and DeFi Ecosystems}
Integrating AI tokens with the broader decentralized finance (DeFi) ecosystem presents new pathways for utility, composability, and value creation. AI models can enhance DeFi by serving as oracles or decision engines for yield optimization, credit scoring, and risk assessment for DeFi, thereby generating demand for AI tokens as payment for such services. Conversely, DeFi can finance AI development through tokenized fundraising, staking mechanisms, or revenue-sharing smart contracts. Additionally, the convergence with NFTs introduces opportunities to tokenize AI models as intellectual property, enabling fractional ownership, royalty flows, and secondary markets tied to model usage. This could support an emerging economy where AI-generated content and services are directly monetized and governed on-chain. 

\subsection{Ethical and Equitable AI Ecosystems}
The future of AI-token networks will increasingly focus on fostering ethical, inclusive, and socially beneficial AI systems. Decentralized architectures have the potential to democratize access to AI [\ref{ref:cihon2020}]. To avoid re-centralization of power, future designs may incorporate mechanisms like quadratic staking, capped rewards, or diversity incentives to ensure broad participation and prevent dominance by large stakeholders or compute-rich actors. Moreover, decentralized governance models could offer a path toward more transparent and accountable AI development, where communities set ethical boundaries and align AI behavior with shared values. Novel frameworks such as "red-team DAOs" [\ref{ref:rikken2023}, \ref{ref:friedler2023}, \ref{ref:feffer2024}] may emerge, rewarding contributors who identify model failures or biases, thereby incentivizing safety and fairness in open-source AI ecosystems. These directions suggest a role for AI tokens not only in technical coordination but also in shaping more equitable and value-aligned AI systems.

In summary, the future of AI-based tokens depends on addressing core limitations through verifiable computation, purpose-built blockchain infrastructures, and sustainable economic models. Emerging technologies such as zero-knowledge proofs and AI-specific consensus mechanisms offer promising pathways to enhance trust, scalability, and composability. Structurally, collaborative training frameworks, interoperable networks, and dynamic incentive systems [\ref{ref:jaiman2022}] may enable the development of more resilient and inclusive ecosystems. If these innovations converge, decentralized AI platforms could support open and globally accessible intelligence development, facilitating new modes of data and model sharing while preserving autonomy and ethical oversight. Realizing this vision will require interdisciplinary collaboration (i.e., integrating advances in cryptography, distributed systems, token economics, and AI safety) as well as proactive engagement with evolving regulatory frameworks [\ref{ref:kumar2021}] to ensure compliance, promote responsible innovation, and align decentralized AI systems with broader societal expectations.

\subsection{Guidelines for Launching a Decentralized AI Token}

Drawing on the patterns observed in our review summarized in Table~\ref{tab:summary_achievement_shortcomings}, which highlights the key strengths and persistent limitations of leading AI-token projects, it becomes evident that many initiatives face common design and adoption challenges. These findings underscore the importance of critically assessing the rationale and structure behind launching a new token especially AI-based token. To support more sustainable and purpose-driven deployments, the following framework offers practical guidelines for evaluating whether, and under what conditions, an AI-based token should be introduced.

\begin{itemize}
\item \textbf{Validate the Need for Decentralization:} Assess whether decentralization is essential to the project’s goals [\ref{ref:hoffman2020}]. If the AI service can operate effectively under a centralized model, the introduction of a token may introduce unnecessary complexity. Tokens should only be deployed where decentralization provides clear benefits, such as enhanced data ownership, censorship resistance, or trustless coordination.

\item \textbf{Establish Product Market Fit:} Before launching a token, ensure that the core AI product addresses a well-defined user need or pain point. Product market fit should be validated through user feedback, prototype testing, or early adoption metrics. Many decentralized AI projects risk failure by attempting to replicate centralized models without sufficient evidence that decentralization improves the user experience or solves a meaningful problem. Tokens should not be introduced merely as a funding mechanism or technical novelty; instead, their design and function must be grounded in a service that delivers clear and differentiated value to its target users. Without product–market fit, token demand will be speculative at best and unsustainable in the long term.

\item \textbf{Define Clear Token Utility:} Specify the token’s function within the platform. Tokens should serve a critical role such as enabling access to AI services, facilitating payments, or supporting governance. A token that lacks unique, indispensable utility risks redundancy and diminished value. Committing to accept only the native token for core services can help anchor its economic relevance.

\item \textbf{Align Incentives and Promote Network Effects:} Design token mechanics that incentivize productive contributions (e.g., data sharing, model development, compute provisioning). The token should help coordinate stakeholder behavior and amplify network growth. However, care must be taken to avoid speculative incentives that detract from the platform’s actual functionality [\ref{ref:catalini2018}].

\item \textbf{Assess Technical Readiness:} Ensure that the AI infrastructure is sufficiently developed to support meaningful token use. Launching a token prematurely before services are usable or integrated with the token can damage credibility. Consider technical factors such as workload distribution (on-chain vs. off-chain) and the feasibility of token integration with core workflows.

\item \textbf{Consider Regulatory and Governance Implications:} Evaluate the legal and governance frameworks surrounding the token. Determine whether the token might be classified as a security or financial instrument in key jurisdictions, and structure the launch to meet decentralization thresholds where applicable. Additionally, prepare for community led governance, including treasury management and protocol upgrades, before transferring control of the ecosystem to token holders.
\end{itemize}

This structured approach helps ensure that token launches within decentralized AI systems are grounded in genuine utility, supported by technical maturity, and guided by responsible governance.

\section{Conclusion}
\label{sec:conclusion}

This paper has critically examined the evolving landscape of AI-based crypto tokens, analyzing their current capabilities, limitations, and future trajectories. Through case studies of projects such as SingularityNET, Ocean Protocol, Fetch.ai (which has now merged into the Artificial Superintelligence Alliance, ASI), Numerai, and Bittensor, we have highlighted how blockchain infrastructure can, in principle, support decentralized AI ecosystems by enabling marketplaces for algorithms and data, crowdsourced prediction systems, and agent-driven economies. However, despite notable technical creativity, most existing platforms remain constrained by fundamental challenges. Chief among these are a heavy reliance on off-chain computation, limited mechanisms for on-chain verifiability, scalability bottlenecks, and weak network effects. As a result, many AI-token ecosystems currently fall short of delivering on the full promise of decentralized, trustless AI services, and in some cases risk functioning primarily as speculative financial instruments rather than engines of genuine innovation.

While our focus has been on AI-token ecosystems, existing research suggests that many of the limitations observed are not unique to this domain. Empirical studies have documented a broader pattern across tokenized ecosystems, where utility tokens often fail to gain sustained adoption and are frequently used more for speculation than for their intended purposes. These findings underscore that the challenges faced by AI-based tokens such as weak product market fit, lack of genuine decentralization, and misaligned incentives may be symptomatic of deeper structural issues in how tokens are designed and deployed within Web3 environments.

Nevertheless, the field is actively evolving. Innovations such as zero-knowledge proofs for machine learning, trusted execution environments, specialized AI blockchains, and federated learning coordinated through smart contracts present promising pathways toward more verifiable, scalable, and composable decentralized AI frameworks. Furthermore, deeper integration with Web3 infrastructures including AI oracles, model NFTs, and agent-governed DAOs may eventually unlock new forms of programmable intelligence and decentralized autonomy.

Critically, the governance of these ecosystems must mature alongside technical developments to ensure alignment with human values, prevent the re-centralization of power, and foster ethical, equitable innovation. Addressing these interdisciplinary challenges will require coordinated advances across cryptography, distributed systems, AI safety, and tokenomics.

In conclusion, while AI-based crypto tokens represent a compelling vision at the intersection of blockchain and artificial intelligence, the current reality remains far from the ideal of decentralized, trustless AI. Whether these ecosystems can overcome their foundational limitations or whether they remain largely an illusion fueled by speculative narratives remains an open question. Future research, technical breakthroughs, and robust governance frameworks will be pivotal in determining whether decentralized AI can move beyond hype toward meaningful, sustainable impact.

\section*{Author Contributions}
Rischan Mafrur: Conceptualization, Methodology, Investigation, Writing Original Draft, Writing Review \& Editing, Supervision.

\section*{Acknowledgments}
The author would like to thank the anonymous reviewers for their constructive and valuable feedback, which significantly improved the quality of this paper.

\section{References}\label{sec12}

\begin{enumerate}
\item \label{ref:kersic2025} Keršić, V., Turkanović, M.: 'A review on building blocks of decentralized artificial intelligence', \emph{ICT Express}, 2025 (Elsevier)

\item \label{ref:defilippi2012} De Filippi, P., McCarthy, S.: 'Cloud computing: Centralization and data sovereignty', \emph{Eur. J. Law Technol.}, 2012, \textbf{3}, (2)

\item \label{ref:zetzsche2020} Zetzsche, D.A., Arner, D.W., Buckley, R.P.: 'Decentralized finance', \emph{J. Financ. Regul.}, 2020, \textbf{6}, (2), pp. 172--203

\item \label{ref:render2025} Render Network: 'About', https://rendernetwork.com/about, accessed 23 April 2025

\item \label{ref:goertzel2017} Goertzel, B., Giacomelli, S., Hanson, D.,~et al.: 'SingularityNET: A decentralized, open market and inter-network for AIs', \emph{Thoughts, Theories Stud. Artif. Intell. Res.}, 2017

\item \label{ref:mcconaghy2022} McConaghy, T.: 'Ocean Protocol: Tools for the Web3 data economy', in \emph{Handbook on Blockchain} (Springer, 2022), pp. 505--539

\item \label{ref:weeks2018} Weeks, M.: 'The evolution and design of digital economies', https://fetch.ai/uploads/Fetch.AI-Economics-white-paper.pdf, 2018, accessed 23 April 2025

\item \label{ref:craib2017} Craib, R., Bradway, G., Dunn, X.,~et al.: 'Numeraire: A cryptographic token for coordinating machine intelligence and preventing overfitting', Retrieved from https://numer.ai/whitepaper.pdf, 2017, accessed 23 April 2025

\item \label{ref:numerai2025} Numerai: 'Leaderboard', https://numer.ai/leaderboard, accessed 27 April 2025

\item \label{ref:rao2020} Rao, Y., Steeves, J., Shaabana, A.,~et al.: 'Bittensor: A peer-to-peer intelligence market', \emph{arXiv preprint arXiv:2003.03917}, 2020

\item \label{ref:ante2024} Ante, L., Demir, E.: 'The ChatGPT effect on AI-themed cryptocurrencies', \emph{Economics \& Business Letters}, 2024, \textbf{13}, (1)

\item \label{ref:saggu2023} Saggu, A., Ante, L.: 'The influence of ChatGPT on artificial intelligence related crypto assets: Evidence from a synthetic control analysis', \emph{Finance Res. Lett.}, 2023, \textbf{55}, pp. 103993

\item \label{ref:tagde2021} Tagde, P., Tagde, S., Bhattacharya, T.,~et al.: 'Blockchain and artificial intelligence technology in e-Health', \emph{Environ. Sci. Pollut. Res.}, 2021, \textbf{28}, pp. 52810--52831

\item \label{ref:femimol2025} Femimol, R., Joseph, L.N.: 'A comprehensive review of blockchain with artificial intelligence integration for enhancing food safety and quality control', \emph{Innov. Food Sci. Emerg. Technol.}, 2025, pp. 104019

\item \label{ref:charles2023} Charles, V., Emrouznejad, A., Gherman, T.: 'A critical analysis of the integration of blockchain and artificial intelligence for supply chain', \emph{Ann. Oper. Res.}, 2023, \textbf{327}, (1), pp. 7--47

\item \label{ref:hosen2022} Hosen, M., Thaker, H.M.T., Subramaniam, V.,~et al.: 'Artificial intelligence (AI), blockchain, and cryptocurrency in finance: Current scenario and future direction', Proc. Int. Conf. Emerg. Technol. Intell. Syst., 2022, pp. 322--332

\item \label{ref:yang2022} Yang, Q., Zhao, Y., Huang, H.,~et al.: 'Fusing blockchain and AI with metaverse: A survey', \emph{IEEE Open J. Comput. Soc.}, 2022, \textbf{3}, pp. 122--136

\item \label{ref:salah2019} Salah, K., Rehman, M.H.U., Nizamuddin, N., Al-Fuqaha, A.: 'Blockchain for AI: Review and open research challenges', \emph{IEEE Access}, 2019, \textbf{7}, pp. 10127--10149

\item \label{ref:marwala2018} Marwala, T., Xing, B.: 'Blockchain and artificial intelligence', \emph{arXiv preprint arXiv:1802.04451}, 2018

\item \label{ref:hussain2021} Hussain, A.A., Al-Turjman, F.: 'Artificial intelligence and blockchain: A review', \emph{Trans. Emerg. Telecommun. Technol.}, 2021, \textbf{32}, (9), e4268

\item \label{ref:buterin2013} Buterin, V.,~et al.: 'Ethereum white paper', GitHub repository, 2013, https://ethereum.org/en/whitepaper/, accessed 23 April 2025

\item \label{ref:michael2018} Michael, J., Cohn, A., Butcher, J.R.: 'Blockchain technology', \emph{The Journal}, 2018, \textbf{1}, (7), pp. 1--11

\item \label{ref:mccormick1985} McCormick, K.: 'AI Technology', \emph{From GI to Z: A Generational Guide to Technology}, 1985

\item \label{ref:zhou2021} Zhou, Z.-H.: \emph{Machine Learning} (Springer Nature, 2021)

\item \label{ref:lecun2015} LeCun, Y., Bengio, Y., Hinton, G.: 'Deep learning', \emph{Nature}, 2015, \textbf{521}, (7553), pp. 436--444

\item \label{ref:ferber1999} Ferber, J., Weiss, G.: \emph{Multi-agent Systems: An Introduction to Distributed Artificial Intelligence} (Addison-Wesley, 1999)

\item \label{ref:adams2023} Adams, H., Salem, M., Zinsmeister, N.,~et al.: 'Uniswap v4 core [draft]', Tech. rep., Uniswap, 2023

\item \label{ref:sushi2025} Sushi: 'Ethereum Swap', https://www.sushi.com/ethereum/swap, accessed 23 April 2025

\item \label{ref:frangella2022} Frangella, E., Herskind, L.: 'Aave v3 technical paper', Technical report, Aave, 2022, https://github.com/aave/aave-v3-core, accessed 23 April 2025

\item \label{ref:leshner2019} Leshner, R., Hayes, G.: 'Compound: The money market protocol', \emph{White Paper}, 2019

\item \label{ref:ellinger2024} Ellinger, E.W., Mini, T., Gregory, R.W., Dietz, A.: 'Decentralized autonomous organization (DAO): The case of MakerDAO', \emph{J. Inf. Technol. Teach. Cases}, 2024, \textbf{14}, (2), pp. 265--272

\item \label{ref:fiedler2023} Fiedler, I., Ante, L.: 'Stablecoins', in \emph{The Emerald Handbook on Cryptoassets: Investment Opportunities and Challenges} (Emerald Publishing Limited, 2023), pp. 93--105

\item \label{ref:raun2023} Raun, C., Estermann, B., Zhou, L.,~et al.: 'Leveraging machine learning for bidding strategies in miner extractable value (MEV) auctions', \emph{Cryptology ePrint Archive}, 2023

\item \label{ref:wang2022} Wang, Y., Zuest, P., Yao, Y.,~et al.: 'Impact and user perception of sandwich attacks in the DeFi ecosystem', Proc. 2022 CHI Conf. on Human Factors in Computing Systems, 2022, pp. 1--15

\item \label{ref:weintraub2022} Weintraub, B., Torres, C.F., Nita-Rotaru, C.,~et al.: 'A flash (bot) in the pan: Measuring maximal extractable value in private pools', Proc. 22nd ACM Internet Measurement Conf., 2022, pp. 458--471

\item \label{ref:arkham2023} Arkham Intelligence: 'Arkham: A Platform for Deanonymizing the Blockchain', White Paper, July 2023, 
\\ https://info.arkm.com/whitepaper, accessed 23 April 2025

\item \label{ref:arkham2023ai} Arkham Intelligence: 'AI Entity Predictions are now live on Arkham!', 
\\ https://info.arkm.com/announcements/ai-entity-predictions-are-now-live-on-arkham, accessed 23 April 2025

\item \label{ref:mafrur2025} Mafrur, R.: 'Blockchain Data Analytics: Review and Challenges', \emph{arXiv preprint arXiv:2503.09165}, 2025

\item \label{ref:calzada2025} Calzada, I.: 'The (Dis) Illusion of the Web3 Decentralization for Global Governance in the Age of GenAI', \emph{Available at SSRN}, 2025

\item \label{ref:barbereau2023} Barbereau, T., Smethurst, R., Papageorgiou, O., Sedlmeir, J., Fridgen, G.: 'Decentralised finance’s timocratic governance: The distribution and exercise of tokenised voting rights', \emph{Technology in Society}, 2023, \textbf{73}, 102251

\item \label{ref:silberholz2021} Silberholz, J., Wu, D.A.: 'Measuring utility and speculation in blockchain tokens', \emph{Available at SSRN}, 2021

\item \label{ref:zimmerman2020} Zimmerman, P.: 'Blockchain structure and cryptocurrency prices', \emph{Bank of England Working Paper}, 2020

\item \label{ref:mei2022} Mei, K., Sockin, M.: 'A theory of speculation in community assets', \emph{Available at SSRN}, 2022

\item \label{ref:coinbase2023} Cubellis, B.: 'Filecoin (FIL): Dissecting storage market incentives', \emph{Coinbase Institutional Research}, 
\\ https://www.coinbase.com/en-au/institutional/research-insights/research/
\\ tokenomics-review/filecoin-fil-dissecting-storage-market-incentives, accessed 20 May 2025

\noindent \item \label{ref:blythman2022} Blythman, R.; Arshath, M.; Smékal, J.; Shaji, H.; Vivona, S.; Dunmore, T.: ‘Libraries, Integrations and Hubs for Decentralized AI using IPFS’, \emph{arXiv}, 2022 (arXiv:2210.16651)

\noindent \item \label{ref:daniel2022} Daniel, E.; Tschorsch, F.: ‘IPFS and Friends: A Qualitative Comparison of Next Generation Peer-to-Peer Data Networks’, \emph{IEEE Commun. Surveys Tuts.}, 2022, 24(1), IEEE (31–52)

\item \label{ref:patelgraph} Patel, A., Green, S., Beylin, E.: 'The Graph as AI infrastructure',  The Graph as AI Infrastructure

\item \label{ref:graph_ai_infra} The Graph Team: 'The Graph as AI Infrastructure', 
\\ https://thegraph.com/blog/the-graph-ai-crypto/, accessed 27 April 2025

\item \label{ref:near2025} NEAR Foundation: 'NEAR Protocol: The Blockchain for AI', https://near.org/, accessed 20 May 2025

\item \label{ref:yakovenko2018} Yakovenko, A.: 'Solana: A new architecture for a high performance blockchain v0.8.13', White paper, 2018

\item \label{ref:py-substrate-interface} Polkadot Developer Hub: 'Python Substrate Interface', 
\\ https://docs.polkadot.com/develop/toolkit/api-libraries/py-substrate-interface/, accessed 23 April 2025

\item \label{ref:kwon2019} Kwon, J., Buchman, E.: 'Cosmos whitepaper', \emph{A Netw. Distrib. Ledgers}, 2019, \textbf{27}, pp. 1--32

\item \label{ref:buchman2016} Buchman, E.: 'Tendermint: Byzantine fault tolerance in the age of blockchains'. PhD thesis, University of Guelph, 2016

\item \label{ref:hoskinson2017} Hoskinson, C.: 'Cardano white paper', 
\\ https://www.cardano.org/en/whiteboard, 2017, accessed 23 April 2025

\item \label{ref:rejuve2023} Goertzel, B.: 'Rejuve Bio: Bringing Neural-Symbolic AI and Cross-Organism Omics Data Together to Cure Aging', White paper, Rejuve Bio, 2023, https://www.rejuve.bio/post/rejuve-bio-bringing-neural-symbolic-ai-and-cross-organism-omics-data-together-to-cure-aging, accessed 23 April 2025

\item \label{ref:nunet2025} NuNet: 'A Global Economy of Decentralized Computing', https://www.nunet.io/, accessed 23 April 2025

\item \label{ref:asi2025} Artificial Superintelligence Alliance: 'Artificial Superintelligence Alliance', https://superintelligence.io/, accessed 23 April 2025

\item \label{ref:chen2018} Chen, Z., Wang, W., Yan, X., Tian, J.: 'Cortex – AI on blockchain', Tech. rep., Cortex Labs Pte. Ltd., Singapore, 2018

\item \label{ref:aiprotocol2023} AI Protocol: 'AI Protocol V3 Whitepaper', 
\\ https://docs.aiprotocol.info/, accessed 23 April 2025

\item \label{ref:deepbrain2023} DeepBrain Chain: 'Artificial Intelligence Computing Platform Driven by Blockchain', White paper, 2023, \\ https://www.deepbrainchain.org/assets/pdf/DeepBrainChainWhitepaper-en.pdf, accessed 23 April 2025

\item \label{ref:story2025} Story Foundation: 'Story: A Peer-to-Peer Intellectual Property Network', White paper, 2025, 
\\ https://www.story.foundation/whitepaper.pdf, accessed 23 April 2025

\item \label{ref:montes2019} Montes, G.A.; Goertzel, B.: ‘Distributed, decentralized, and democratized artificial intelligence’, \emph{Technol. Forecast. Soc. Change}, 2019, 141, Elsevier (354–358) 

\item \label{ref:huggingface2025} Hugging Face: 'The AI community building the future', https://huggingface.co/, accessed 23 April 2025

\item \label{ref:kaggle2025} Kaggle: 'Your Machine Learning and Data Science Community', https://www.kaggle.com/, accessed 23 April 2025

\item \label{ref:awsai2025} Amazon Web Services: 'Artificial Intelligence on AWS', https://aws.amazon.com/ai/, accessed 23 April 2025

\item \label{ref:azureai2025} Microsoft Azure: 'Artificial Intelligence on Azure', 
\\ https://azure.microsoft.com/en-us/solutions/ai/, accessed 23 April 2025

\item \label{ref:googleai2025} Google: 'Google AI – How we're making AI helpful for everyone', https://ai.google/, accessed 23 April 2025

\item \label{ref:voneschenbach2021} Von Eschenbach, W.J.: 'Transparency and the black box problem: Why we do not trust AI', \emph{Philos. Technol.}, 2021, \textbf{34}, (4), pp. 1607--1622

\item \label{ref:marshall2024} Marshall, T.: 'Decentralised artificial intelligence', 
\\ \emph{CMS Law}, https://cms.law/en/gbr/publication/decentralised-artificial-intelligence, accessed 23 April 2025

\item \label{ref:jacovi2021} Jacovi, A., Marasović, A., Miller, T., Goldberg, Y.: 'Formalizing trust in artificial intelligence: Prerequisites, causes and goals of human trust in AI', Proc. 2021 ACM Conf. on Fairness, Accountability, and Transparency, 2021, pp. 624--635

\item \label{ref:capponi2024} Capponi, A., Jia, R., Yu, S.: 'Price discovery on decentralized exchanges', Available at SSRN 4236993, 2024

\item \label{ref:wang2022} Wang, Y., Zuest, P., Yao, Y.,~et al.: 'Impact and user perception of sandwich attacks in the DeFi ecosystem', Proc. 2022 CHI Conf. on Human Factors in Computing Systems, 2022, pp. 1--15

\item \label{ref:voskobojnikov2021} Voskobojnikov, A., Wiese, O., Mehrabi Koushki, M.,~et al.: 'The U in crypto stands for usable: An empirical study of user experience with mobile cryptocurrency wallets', Proc. 2021 CHI Conf. on Human Factors in Computing Systems, 2021, pp. 1--14

\item \label{ref:kwon2019bitcoin} Kwon, Y., Kim, H., Shin, J., Kim, Y.: 'Bitcoin vs. Bitcoin Cash: Coexistence or downfall of Bitcoin Cash?', Proc. 2019 IEEE Symp. on Security and Privacy (SP), 2019, pp. 935--951

\item \label{ref:kiffer2017} Kiffer, L., Levin, D., Mislove, A.: 'Stick a fork in it: Analyzing the Ethereum network partition', Proc. 16th ACM Workshop on Hot Topics in Networks (HotNets), 2017, pp. 94--100

\item \label{ref:clarke2019} Clarke, R.: 'Regulatory alternatives for AI', \emph{Comput. Law Secur. Rev.}, 2019, \textbf{35}, (4), pp. 398--409

\item \label{ref:uzougbo2024} Uzougbo, N.S., Ikegwu, C.G., Adewusi, A.O.: 'Regulatory frameworks for decentralized finance (DeFi): Challenges and opportunities', \emph{GSC Adv. Res. Rev.}, 2024, \textbf{19}, (2), pp. 116--129

\item \label{ref:openai2025} OpenAI: 'OpenAI – Advancing Digital Intelligence', 
\\ https://openai.com/, accessed 23 April 2025

\item \label{ref:bartoletti2025} Bartoletti, M., Benetollo, L., Bugliesi, M.,~et al.: 'Smart contract languages: A comparative analysis', \emph{Future Gener. Comput. Syst.}, 2025, \textbf{164}, pp. 107563

\item \label{ref:chen2024zkml} Chen, B.-J., Waiwitlikhit, S., Stoica, I., Kang, D.: 'zkML: An optimizing system for ML inference in zero-knowledge proofs', Proc. 19th Eur. Conf. on Computer Systems, 2024, pp. 560--574

\item \label{ref:abbaszadeh2024} Abbaszadeh, K., Pappas, C., Katz, J., Papadopoulos, D.: 'Zero-knowledge proofs of training for deep neural networks', Proc. ACM SIGSAC Conf. on Computer and Communications Security, 2024, pp. 4316--4330

\item \label{ref:kersic2024} Keršič, V., Karakatić, S., Turkanović, M.: 'On-chain zero-knowledge machine learning: An overview and comparison', \emph{Journal of King Saud University - Computer and Information Sciences}, 2024, 102207

\item \label{ref:xing2025} Xing, Z., Zhang, Z., Zhang, Z.,~et al.: 'Zero-Knowledge Proof-Based Verifiable Decentralized Machine Learning in Communication Network: A Comprehensive Survey', \emph{IEEE Communications Surveys \& Tutorials}, 2025

\item \label{ref:breidenbach2021} Breidenbach, L., Cachin, C., Chan, B.,~et al.: 'Chainlink 2.0: Next steps in the evolution of decentralized oracle networks', Chainlink Labs, 2021, \textbf{1}, pp. 1--136

\item \label{ref:chainlink2024} Chainlink: 'The Intersection Between AI Models and Oracles', Chainlink Blog, 4 July 2024, https://blog.chain.link/oracle-networks-ai/, accessed 23 April 2025

\item \label{ref:sabt2015} Sabt, M., Achemlal, M., Bouabdallah, A.: 'Trusted execution environment: What it is, and what it is not', Proc. 2015 IEEE Trustcom/BigDataSE/Ispa, 2015, \textbf{1}, pp. 57--64

\item \label{ref:ball2017} Ball, M., Rosen, A., Sabin, M., Vasudevan, P.N.: 'Proofs of useful work', \emph{Cryptology ePrint Archive}, 2017

\item \label{ref:lihu2020} Lihu, A., Du, J., Barjaktarevic, I.,~et al.: 'A proof of useful work for artificial intelligence on the blockchain', \emph{arXiv preprint arXiv:2001.09244}, 2020

\item \label{ref:baldominos2019} Baldominos, A., Saez, Y.: 'Coin.AI: A proof-of-useful-work scheme for blockchain-based distributed deep learning', \emph{Entropy}, 2019, \textbf{21}, (8), 723

\item \label{ref:huang2024} Huang, C., Song, R., Gao, S.,~et al.: 'Data availability and decentralization: New techniques for zk-rollups in Layer 2 blockchain networks', \emph{arXiv preprint arXiv:2403.10828}, 2024

\item \label{ref:song2024} Song, H., Qu, Z., Wei, Y.: 'Advancing blockchain scalability: An introduction to layer 1 and layer 2 solutions', Proc. 2024 IEEE 2nd Int. Conf. on Sensors, Electronics and Computer Engineering (ICSECE), 2024, pp. 71--76

\item \label{ref:tortola2024} Tortola, D., Lisi, A., Mori, P., Ricci, L.: 'Tethering Layer 2 solutions to the blockchain: A survey on proving schemes', \emph{Comput. Commun.}, 2024

\item \label{ref:sguanci2021} Sguanci, C., Spatafora, R., Vergani, A.M.: 'Layer 2 blockchain scaling: A survey', \emph{arXiv preprint arXiv:2107.10881}, 2021

\item \label{ref:cong2024} Cong, M., Yuen, T.H., Yiu, S.-M.: 'zkmatrix: Batched short proof for committed matrix multiplication', Proc. 19th ACM Asia Conf. on Computer and Communications Security, 2024, pp. 289--305

\item \label{ref:zawistowski2016} Zawistowski, J., Janiuk, P., Regulski, A.: 'The Golem project – crowdfunding whitepaper', Golem, 2016, \textbf{28}

\item \label{ref:iexec2025} iExec: 'Decentralized cloud computing platform', 
\\ https://www.iex.ec/, accessed 28 April 2025

\item \label{ref:qu2022} Qu, Y., Uddin, M.P., Gan, C.,~et al.: 'Blockchain-enabled federated learning: A survey', \emph{ACM Comput. Surv.}, 2022, \textbf{55}, (4), pp. 1--35

\item \label{ref:toyoda2020} Toyoda, K., Zhao, J., Zhang, A.N.S., Mathiopoulos, P.T.: 'Blockchain-enabled federated learning with mechanism design', \emph{IEEE Access}, 2020, \textbf{8}, pp. 219744--219756

\item \label{ref:qu2020} Qu, Y., Gao, L., Luan, T.H.,~et al.: 'Decentralized privacy using blockchain-enabled federated learning in fog computing', \emph{IEEE Internet Things J.}, 2020, \textbf{7}, (6), pp. 5171--5183

\noindent \item \label{ref:ren2024} Ren, S.; Kim, E.; Lee, C.: ‘A scalable blockchain-enabled federated learning architecture for edge computing’, \emph{PLOS ONE}, 2024, 19(8), Public Library of Science (e0308991) 

\item \label{ref:shayan2021} Shayan, M.; Fung, C.; Yoon, C.J.M.; Beschastnikh, I.: ‘Biscotti: A Blockchain System for Private and Secure Federated Learning’, \emph{IEEE Trans. Parallel Distrib. Syst.}, 2021, 32(7), IEEE (1513–1525) 

\noindent \item \label{ref:mothukuri2021} Mothukuri, V.; Parizi, R.M.; Pouriyeh, S.; Dehghantanha, A.; Choo, K.-K.R.: ‘FabricFL: Blockchain-in-the-Loop Federated Learning for Trusted Decentralized Systems’, \emph{IEEE Systems Journal}, 2021, 16(3), IEEE (3711–3722)

\noindent \item \label{ref:weng2019} Weng, J.; Zhang, Y.; Li, M.; Zhang, J.; Luo, W.; Weng, J.: ‘DeepChain: Auditable and Privacy-Preserving Deep Learning with Blockchain-Based Incentive’, \emph{IEEE Trans. Dependable Secure Comput.}, 2019, 16(5), IEEE 

\item \label{ref:potts2021} Potts, J., MacDonald, T.: 'New data city: The future of the digital CBD as a data pool on a DAO', Available at SSRN 3892009, 2021

\item \label{ref:thomas2024} Thomas, A.J.: 'DAOs, cooperatives and the educational principle', \emph{Law Ethics Technol.}, 2024, \textbf{1}, pp. 1--30

\item \label{ref:petreski2024} Petreski, D., Cheong, M.: 'Data cooperatives: A conceptual review', 2024

\item \label{ref:zichichi2022} Zichichi, M., Ferretti, S., Rodríguez-Doncel, V.: 'Decentralized personal data marketplaces: How participation in a DAO can support the production of citizen-generated data', \emph{Sensors}, 2022, \textbf{22}, (16), pp. 6260

\noindent \item \label{ref:jaiman2022} Jaiman, V.; Pernice, L.; Urovi, V.: ‘User incentives for blockchain-based data sharing platforms’, \emph{PLOS ONE}, 2022, 17(4), Public Library of Science (e0266624) (

\item \label{ref:nguyen2021} Nguyen, L.D., Pandey, S.R., Soret, B.,~et al.: 'A marketplace for trading AI models based on blockchain and incentives for IoT data', \emph{arXiv preprint arXiv:2112.02870}, 2021

\item \label{ref:zhang2024} Zhang, Z., Rao, Y., Xiao, H.,~et al.: 'Proof of quality: A costless paradigm for trustless generative AI model inference on blockchains', \emph{arXiv preprint arXiv:2405.17934}, 2024

\item \label{ref:ayaz2020} Ayaz, F., Sheng, Z., Tian, D., Guan, Y.L.: 'A proof-of-quality-factor (PoQF)-based blockchain and edge computing for vehicular message dissemination', \emph{IEEE Internet Things J.}, 2020, \textbf{8}, (4), pp. 2468--2482

\item \label{ref:ocean_token_model} Ocean Protocol Team: 'Ocean Token Model', 
\\ https://blog.oceanprotocol.com/ocean-token-model-3e4e7af210f9, accessed 27 April 2025

\noindent \item \label{ref:shafay2023} Shafay, J.; Nadir, A.; Yuan, Y.; Hassan, A.U.; Shah, Y.; Anwar, M.W.; Koubaa, A.: ‘Blockchain for deep learning: review and open challenges’, \emph{Cluster Comput.}, 2023, 26, Springer (197–221) 

\item \label{ref:das2024} Das, P.; Singh, M.; Verma, K.K.: ‘Blockchain-Enabled Deep Learning Approach to Improve Healthcare System’, \emph{J. Multimedia Inf. Syst.}, 2024, 11(1), Korea Multimedia Soc. (9–16) 

\noindent \item \label{ref:kumar2021} Kumar, A.; Finley, B.; Braud, T.; Tarkoma, S.; Hui, P.: ‘Sketching an AI Marketplace: Technological, Economic, and Regulatory Aspects’, \emph{IEEE Access}, 2021, 9, IEEE 

\item \label{ref:feffer2024} Feffer, M., Sinha, A., Deng, W.H.,~et al.: 'Red-teaming for generative AI: Silver bullet or security theater?', Proc. AAAI/ACM Conf. on AI, Ethics, and Society, 2024, \textbf{7}, pp. 421--437

\item \label{ref:friedler2023} Friedler, S., Singh, R., Blili-Hamelin, B.,~et al.: 'AI red-teaming is not a one-stop solution to AI harms', 2023

\item \label{ref:cihon2020} Cihon, P., Maas, M.M., Kemp, L.: 'Should artificial intelligence governance be centralised? Design lessons from history', Proc. AAAI/ACM Conf. on AI, Ethics, and Society, 2020, pp. 228--234

\item \label{ref:rikken2023} Rikken, O.K., Janssen, M.F.W.H.A., Roosenboom-Kwee, Z.: 'The ins and outs of decentralized autonomous organizations (DAOs): Unraveling the definitions, characteristics, and emerging developments of DAOs', \emph{Blockchain: Research and Applications}, 2023, \textbf{4}, (3), 100143

\item \label{ref:hoffman2020} Hoffman, M.R., Ibáñez, L.-D., Simperl, E.: 'Toward a formal scholarly understanding of blockchain-mediated decentralization: A systematic review and a framework', \emph{Frontiers in Blockchain}, 2020, \textbf{3}, 35

\item \label{ref:catalini2018} Catalini, C., Gans, J.S.: 'Initial coin offerings and the value of crypto tokens', \emph{National Bureau of Economic Research Working Paper}, 2018

\end{enumerate}

\end{document}